\documentclass[10pt,aps,prb,twocolumn, nofootinbib]{revtex4-1}

\usepackage{amssymb,amsmath}
\usepackage{subfig}
\usepackage{graphicx}
\usepackage{color}
\usepackage{hyperref}
\usepackage{listings}
\usepackage[font=small, labelfont=bf, justification=justified, format=plain]{caption} 
\hypersetup{
    bookmarks=true,         
    unicode=false,          
    pdftoolbar=true,        
    pdfmenubar=true,        
    pdffitwindow=false,     
    pdfstartview={FitH},    
    pdfauthor={Eitan Dvorquez},     
    colorlinks=true,       
    linkcolor=blue,          
    citecolor=red,        
}
\graphicspath{{figures/}}

\newcommand{\specialcell}[2][c]{%
  \begin{tabular}[#1]{@{}c@{}}#2\end{tabular}}
\newcommand\ddfrac[2]{\frac{\displaystyle #1}{\displaystyle #2}}

\begin{document}
\definecolor{dkgreen}{rgb}{0,0.6,0}
\definecolor{gray}{rgb}{0.5,0.5,0.5}
\definecolor{mauve}{rgb}{0.58,0,0.82}

\lstset{frame=tb,
  	language=Matlab,
  	aboveskip=3mm,
  	belowskip=3mm,
  	showstringspaces=false,
  	columns=flexible,
  	basicstyle={\small\ttfamily},
  	numbers=none,
  	numberstyle=\tiny\color{gray},
 	keywordstyle=\color{blue},
	commentstyle=\color{dkgreen},
  	stringstyle=\color{mauve},
  	breaklines=true,
  	breakatwhitespace=true
  	tabsize=3
}

\title{Mu-Metal Enhancement of Effects in Electromagnetic Fields Over Single
Emitters Near Topological Insulators}
\author{Eitan Dvorquez${}^a$, Benjamín Pavez${}^a$, Qiang Sun${}^{b}$, Felipe Pinto${}^a$, Andrew D. Greentree${}^{b}$, Brant C. Gibson${}^{b}$, and Jerónimo R. Maze${}^a$}

\affiliation{${}^a$Institute of Physics, Pontificia Universidad Católica de Chile, Santiago, Chile.}
\affiliation{\resizebox{0.8\textwidth}{!}{${}^b$ARC Centre of Excellence for Nanoscale Biophotonics, RMIT University, Melbourne, VIC 3001,
Australia.}}
\date{\today}

\begin{abstract}
We focus on the transmission and reflection coefficients of light in systems involving of topological insulators (TI). Due to the electro-magnetic coupling in TIs, new mixing coefficients emerge leading to new components of the electromagnetic fields of propagating waves. 
We have discovered a simple heterostructure that provides a 100-fold enhancement of the mixing coefficients for TI materials. Such effect increases with the TI's wave impedance. We also predict a transverse deviation of the Poynting vector due to these mixed coefficients contributing to the radiative electromagnetic field of an electric dipole. Given an optimal configuration of the dipole-TI system, this deviation could amount to $0.18\%$ of the Poynting vector due to emission near not topological materials, making this effect detectable. 

\end{abstract}

\maketitle

\section{\label{sec:one}Introduction }
Topological insulators (TIs) are materials with non-trivial topological order. These materials present an insulating bulk, while still having conductive edge states on their surface \cite{Qi_Zhang, Hasan_Moore2011, Yoichi2013}. These properties allow interesting electronic effects such as the quantum spin Hall effect in 2D TIs, which is a quantized Hall current that is present despite the absence of an external magnetic field.

TIs exhibit unusual optical properties, such as the coupling between the electric and magnetic fields, which produces magnetic monopole-like behavior when a point charge is near a TI's surface \cite{point_charge}, varying Hall conductivity \cite{Essin_2009}, and the magneto-electric effect\cite{jero+martin}. Remarkably, the magneto-electric effect that TIs exhibits has been measured to respond faster than multiferroic's magneto-electric effect\cite{Kitagawa2010, multiferroic} and does so without material fatigue\cite{Moore2010}.

In recent years, these properties have been at the center of the development of topological devices such as nanowires\cite{Legg2022, PhysRevResearch.4.013087, Mnning2021}, memory storage\cite{Wu2021}, nanoparticles\cite{CastroEnrquez2022,Lin2015}, among others. Application of these devices have been proposed for optoelectronics, spintronics and quantum computing\cite{He2019,Pandey2021,Breunig2021}. To make such devices, it is crucial to develop reliable methods to characterize TIs.

Efforts to characterize the optical effects exhibited by TIs usually seek to measure the Faraday rotation of a transmitted ray, as it should be quantized by the fine structure constant $\alpha$\cite{Dziom2017, JC2017}. However, this effect is usually very small (1-10 mrad), and the conditions for it to be purely topological are such that it requires a very strong magnetic field (on the Tesla range), low temperatures, and low-frequency emissions (THz range) \cite{Crosse2017, AShuvaev:2022, Tse_2010, Tse_2011}. The Faraday effect can be enhanced by a wide range of material configurations. Composite TI multi-layered media materials\cite{Crosse_2016} and defects on the TI's surface\cite{GhasempourArdakani:21} are examples of such systems. The geometry of the system can also have an impact on the enhancement of these effects\cite{MartnRuiz2016, Ge2014, HassaniGangaraj2020}. It is also possible to observe the rotation on semi-magnetic TIs, without a magnetic field\cite{Mogi2022}. 

\begin{figure}[!htp]
    \centering
    \includegraphics[width=0.9\linewidth]{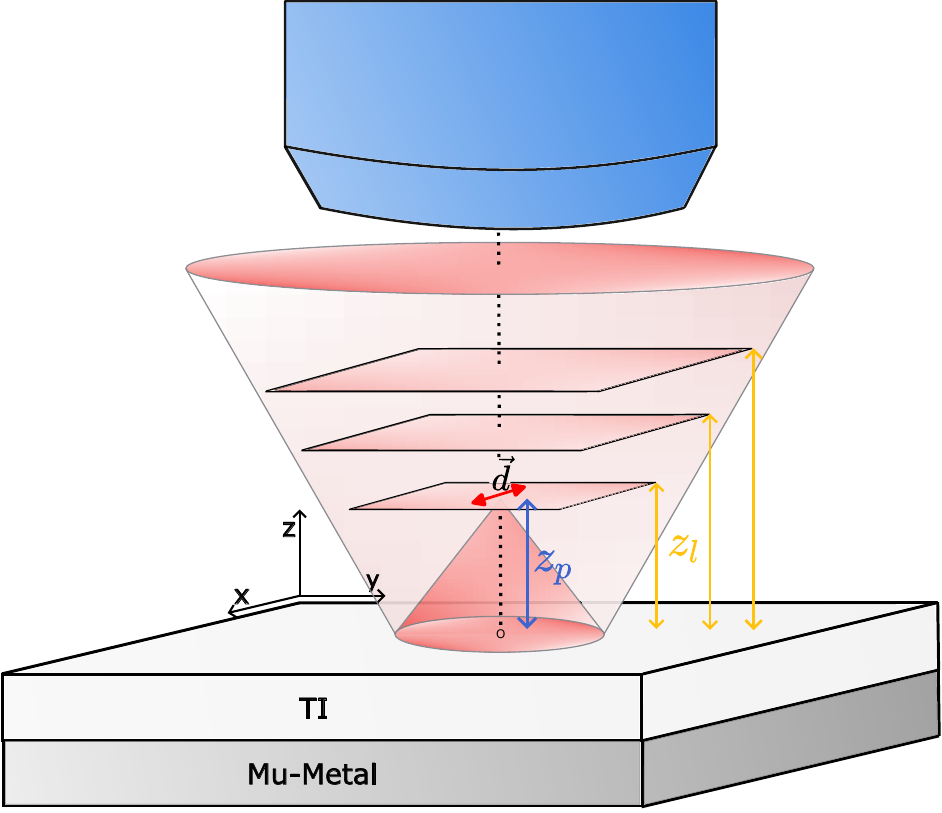}
    \caption{Sketch of the studied dipole-TI-Mu metal system for an $\hat{x}$ oriented dipole $\vec{d}$, where $z_p$ is the distance of the dipole relative to the TI's surface, and $z_l$ is the distance from the surface to the observation plane. }
    \label{fig:Sketch_lens_dipole_system}
\end{figure}

Here we show an optical method to characterize TIs. We want to find new, fast, and spatially resolved techniques to characterize TIs optically. With this in mind, we will start by setting the foundations for the optical characterization of a TI's planar surface via the Axion Lagrangian, remarking on the fundamental aspects of the derivation for new boundary conditions, Fresnel coefficients, and the electric field's Green's function. Then, we will study the inclusion of a Mu-Metal layer into the system to enhance these topological effects, in a setup like the one in figure \ref{fig:Sketch_lens_dipole_system}.

\section{Theoretical Background}
\subsection{\label{sec:oneCond}Axion Formalism for Topological Insulators}
To describe the effect of an interface over any incident emission, we start by obtaining the Maxwell equations and boundary conditions.

Due to this electromagnetic coupling, we can describe these materials via axion theory\cite{MartnRuiz2021}, adding an extra term to the usual Maxwell Lagrangian, such that $\mathcal{L} = \mathcal{L}_{M} + \mathcal{L}_{\Theta}$, where
\begin{align}
    \mathcal{L}_{M} =& \frac{1}{8\pi}\left(\epsilon |\bf{E}|^2-\frac{1}{\mu}|\bf{B}|^2\right),\\
    \mathcal{L}_{\Theta} =& \frac{\alpha}{4\pi^2}\frac{\Theta(\mathbf{r})}{\mu_0 c^2} \mathbf{E}\cdot\mathbf{B},
\end{align}

\noindent where $\bf{E}$ and $\bf{B}$ are the electric and magnetic fields, respectively, $\mu_0$ is the vacuum's magnetic permeability, and $c$ is the speed of light. $\alpha$ is the fine structure constant and $\Theta(\bf{r})$ is the axionic coupling parameter as a function of position in space $\bf{r}$, which is 0 for a non-topological material and an odd multiple of $\pi$ for TIs\cite{Qi_Zhang}. 
The Euler-Lagrange equation outputs a new set of modified time-dependent Maxwell equations; 
\begin{gather}
    \nabla \cdot \mathbf{E} = \mu_0 c^2 \rho - c\frac{\alpha }{4\pi^2}\nabla\Theta\cdot\mathbf{B}, \label{eq:max_1}\\
    \nabla\times \textbf{B} =\mu_0\textbf{J} + \frac{1}{c^2}\frac{\partial \textbf{E}}{\partial t} + \frac{1}{c}\frac{\alpha}{4\pi^2}\nabla\Theta\times\mathbf{E} + \frac{1}{c}\frac{\partial \Theta}{\partial t}\mathbf{B}, \label{eq:max_2}\\
    \nabla \times \textbf{E}= -\tfrac{\partial \textbf{B}}{\partial t}, \label{eq:max_3}\\
    \nabla \cdot \textbf{B}=0, \label{eq:max_4}
\end{gather}
and, therefore, new terms on the boundary conditions to the half-plane space problem of an interface Air-TI, are directly dependent on the difference between the axion parameters of the two materials. 

Taking advantage of the linear relations between $\mathbf{E}$ and $\mathbf{B}$ upon converting the time $t$ to frequency $\omega$ via Fourier Transform, we define the $\mathbf{D}$ and $\mathbf{H}$ fields as
\begin{align}
    \mathbf{D}(\mathbf{r},\omega) &= \epsilon_0\epsilon_\mathrm{r}(\mathbf{r},\omega)\mathbf{E}(\mathbf{r},\omega)+\frac{\alpha}{\pi}\frac{\Theta(\mathbf{r})}{\mu_0 c}\mathbf{B}(\mathbf{r},\omega),\label{eq:D_topo}\\
    \mathbf{H}(\mathbf{r},\omega)&=\frac{1}{\mu_0\mu_\mathrm{r}(\mathbf{r},\omega)}\mathbf{B}(\mathbf{r},\omega)-\frac{\alpha}{\pi}\frac{\Theta(\mathbf{r})}{\mu_0 c}\mathbf{E}(\mathbf{r},\omega).\label{eq:H_topo}
\end{align}
where $\epsilon_0$ is the vacuum's electric permittivity, and $\epsilon_\mathrm{r}(\mathbf{r},\omega)$ and $\mu_\mathrm{r}(\mathbf{r},\omega)$ are the relative electric permittivity and magnetic permeability of the TI, respectively. With these definitions, we can retain the well-known boundary equations $\hat{z}\times \mathbf{E}_1 = \hat{z}\times \mathbf{E}_2$ and $\hat{z}\times \mathbf{H}_1 = \hat{z}\times \mathbf{H}_2$, where $\hat{z}$ is the direction on the axis perpendicular to the TI's plane and the subscripts $1$ and $2$ indicate fields over and under the TI's surface, respectively. Also, $ \mathbf{D}$ and $ \mathbf{H}$ follow the usual Maxwell equations.

The new Helmholtz wave equation is obtained by combining the modified Maxwell equations:
\begin{equation}\label{eq:helmholtz_topo}\textstyle{}
    \nabla \times \nabla \times \mathbf{E} - \frac{\omega^2}{c^2}\mathbf{E} - i\omega\mu_0\left(\frac{\alpha}{4\pi^2 \mu_0c}\right)\nabla\Theta\times\mathbf{E} = -i\omega\mu_0\mathbf{J},
\end{equation}
\noindent where $\mathbf{J}$ stands for the usual current, constituted by the external free current and the current generated by the non-topological magnetodielectric responses of the material. 

The extra term in equation (\ref{eq:helmholtz_topo}) can be understood as an extra superficial current due to the topological properties of the material, given by the coupling parameter $\Theta(\mathbf{r})$. This current is exclusively located on the surface of the interface due to the step function like the behavior of $\Theta(\mathbf{r})$, which implies that $\nabla\Theta(\mathbf{r})$ behaves as a Dirac delta on the interface of the materials. 

\subsection{\label{sec:twoCond} Two-Layered Half-Plane System}
To characterize the reflected or transmitted waves, we explore the Fresnel coefficients that follow these new boundary conditions. Comparing these new coefficients with the regular ones for a non-topological material gives us insight into the effect we seek to measure. We start by considering a planar wave ansatz. 

When considering an incident planar wave, we must take into account its polarization; Transverse Electric (TE) or Transverse Magnetic (TM), referring to which field, $\mathbf{E}$ or $\mathbf{H}$, is initially pointing in the $\hat{y}$ axis. For magnetodielectric media, reflected and transmitted waves maintain the initial axis of their respective fields. On the other hand, due to the electromagnetic coupling, the waves interacting with TIs present new fields components on all axes. For example, given a TE polarized wave, with $\mathbf{E}=E_0 \hat{y}$, we should expect the reflected wave to present a resulting electric field component not only on the $\hat{y}$ axis, but on the $\hat{x}$ and $\hat{z}$ axis as well.  

Using a planar wave ansatz for the electric and magnetic field of initially TE and TM polarized waves, and combining them with the new boundary conditions previously discussed yields the following new reflection and transmission matrices\cite{TSB-TI},

\begin{widetext}
\begin{align}
    \mathbf{R}&=\begin{pmatrix}R_\textit{TE,TE} & R_\textit{TE,TM}\\ R_\textit{TM,TE} & R_\textit{TM,TM}\end{pmatrix} =\frac{1}{\gamma}\left[\begin{matrix}(\mu_2k_{z1}-\mu_1k_{z2})\Omega_\epsilon-k_{z1}k_{z2}\Delta^2&-2\mu_2n_1k_{z1}k_{z2}\Delta\\-2\mu_2n_1k_{z1}k_{z2}\Delta&(\epsilon_2k_{z1}-\epsilon_1k_{z2})\Omega_\mu-k_{z1}k_{z2}\Delta^2\end{matrix}\right],\\
    \mathbf{T}&=\begin{pmatrix}T_\textit{TE,TE} & T_\textit{TE,TM}\\ T_\textit{TM,TE} & T_\textit{TM,TM}\end{pmatrix} =\frac{2k_{z1}}{\gamma}\left[\begin{matrix}\mu_2\Omega_\epsilon & -\mu_2n_1k_{z2}\Delta\\\mu_2n_2k_{z1}\Delta & \tfrac{n_2}{n_1}\epsilon_1\Omega_\mu\end{matrix}\right],
\end{align}
\end{widetext}
\normalsize

\noindent where $\epsilon_i$ and $\mu_i$ are the relative electric permittivity  and magnetic permeability, respectively, of the $i$-layer. $\mathbf{k}_i=(k_{xi},k_{yi},k_{zi})$ is the wave vector in each layer, $n_i=\sqrt{\mu_i\epsilon_i}$ is the refractive index of the corresponding layer, $\Delta=\alpha\mu_1\mu_2(\Theta_2-\Theta_1)/\pi$, $\Omega_\mu = \mu_1\mu_2(k_{z1}\mu_2+k_{z2}\mu_1)$, $\Omega_\epsilon = \mu_1\mu_2(k_{z1}\epsilon_2+k_{z2}\epsilon_1)$ and $\gamma = \frac{\Omega_\mu\Omega_\epsilon}{\mu_1\mu_2}+k_{z1}k_{z2}\Delta^2$. 

It is worth remarking that, when $\Theta_2-\Theta_1\rightarrow 0$ the mixed coupling coefficients $R_\textit{TM,TE}$, $R_\textit{TE,TM}$, $T_\textit{TE,TM}$, $T_\textit{TM,TE}\rightarrow0$ and $R_\textit{TE,TE}$, $R_\textit{TM,TM}$, $T_\textit{TE,TE}$, and $T_\textit{TM,TM}$ reduce to the conventional coefficients for magnetodielectric media \cite{weng_cho_waves, nano_optics_2012}. Further limit cases are found in table \ref{tab:coeff_limits}.

\begin{table*}[!htp]
    \centering
    \begin{tabular}{|c|c|c|c|c|c|c|}
        \hline
        & \specialcell{Air/Perfect\\Conductor} & Air/Mu-Metal & Mu-Metal/Air & \specialcell{TI/Perfect\\Conductor} & TI/Mu-Metal & Mu-Metal/TI\\[5ex]
        \hline &&&&&&\\
         \specialcell{Reflection\\Matrix} & $\left[\begin{matrix}-1&0\\0&1\end{matrix}\right]$ & $\left[\begin{matrix}1&0\\0&-1\end{matrix}\right]$ & $\left[\begin{matrix}-1&0\\0&1\end{matrix}\right]$ & $\left[\begin{matrix}-1&0\\0&1\end{matrix}\right]$ & $\left[\begin{matrix}1&0\\0&-1\end{matrix}\right]$ & $\left[\begin{matrix}-1&0\\0&1\end{matrix}\right]$ \\[5ex]
         \hline &&&&&&\\
         \specialcell{Transmission\\Matrix} & $\left[\begin{matrix}0&0\\0&0\end{matrix}\right]$ & $\left[\begin{matrix}2&0\\0&2\frac{\cos{\phi_r}}{\cos{\phi_t}}\end{matrix}\right]$ & $\left[\begin{matrix}0&0\\0&0\end{matrix}\right]$ & $\left[\begin{matrix}0&0\\0&0\end{matrix}\right]$ & $\ddfrac{1}{\epsilon_1+\tilde{\alpha}^2\mu_1}\left[\begin{matrix}2\epsilon_1&\frac{n_2k_{z1}}{k_{z2}}2\tilde{\alpha}\\-2n_1\tilde{\alpha}&\frac{n_2k_{z1}}{n_1k_{z2}}2\epsilon_1\end{matrix}\right]$ & $\left[\begin{matrix}0&0\\0&0\end{matrix}\right]$ \\[5ex]
         \hline
    \end{tabular}
    \caption{Limit case results of the Fresnel matrices for different 2 layered-Systems involving Perfect Conductors ($\epsilon \rightarrow \infty$) and Mu-Metals ($\mu \rightarrow \infty$). The perfect conductor retains its reflecting behavior when paired with TIs. The Mu-metal and TI pairing retains the Mu-Metal shielding phenomena, irrespective of the new off-diagonal coefficients in the TI/Mu-Metal configuration, given that the transmission coefficients of the Mu-Metal/TI configuration are null.}
    \label{tab:coeff_limits}
\end{table*}

The Fresnel coefficients are given by 
\begin{equation}\label{eq:real_coefficients}
    r_{ij} = \left| R_{ij} \right|^2, \qquad t_{ij} = \frac{k_{z2}}{k_{z1}}\frac{\mu_1}{\mu_2}\left| T_{ij} \right|^2.
\end{equation}
New mixed Fresnel coefficients, derived from the new off-diagonal terms on the $\mathbf{R}$ and $\mathbf{T}$ matrices, are about $10^{-5}$ times smaller compared to the usual non-mixed ones\cite{TSB-TI} because they scaled as $\alpha^2$. This indicates that any effects from these coefficients are very small 
with respect to the usual fields. 

The reflected mixed coefficients tend to increase with the TI's impedance relative to the air. However, since the impedance is an intrinsic property of the material and the increase is relatively minor, we will search for another way to maximize these coefficients. Topologically non-trivial heterostructures have been widely proposed to reach said goal, primarily because proximity coupling introduces magnetism to the TI without modifying the structure of the material, both enabling and enhancing topological response\cite{Liu2021, Mogi2017, An2020, Zuo2013, Crosse_2016}.  Similarly, we propose the addition of a single Mu-Metal layer below the TI.

\subsection{Electric Green's Function}
We define the Green's function such that it satisfies
\begin{equation}\label{eq:E_and_G_with_J}
    \mathbf{E}(\mathbf{r},\omega)=i\omega\mu_0\int d^3r' \mathbf{G}(\mathbf{r},\mathbf{r}',\omega)\cdot\mathbf{J}(\mathbf{r}'),
\end{equation}

\noindent then, knowing the behavior of $\mathbf{G}$ gives us the behavior of $\mathbf{E}$.
The Green's function also satisfies the following Helmholtz equation,
\begin{equation}
    \nabla\times\frac{1}{\mu_\mathrm{r}(\mathbf{r},\omega)}\nabla\times\mathbf{G}(\mathbf{r},\mathbf{r}',\omega) - \frac{\omega^2}{c^2}\epsilon_\mathrm{r}(\mathbf{r},\omega)\mathbf{G}(\mathbf{r},\mathbf{r}',\omega)=\delta(\mathbf{r}-\mathbf{r}').
\end{equation}
This equation is equal to the one for a traditional magnetodielectric, as we are interested in the Green's function on the bulk of the upper layer and not on the interface. 

The corresponding Green's function, in its singularity extracted form, for non-diagonal Fresnel matrices, is\cite{TSB-TI}
\begin{align}\label{eq:general_gf}
    \mathbf{G}(\mathbf{r},\mathbf{r}',\omega) =& \frac{i}{8\pi^2}\int d^2k_p\,\mu_{\mathrm{r}}(\omega)\times\nonumber \\
    &\left[ \frac{\mathbf{C}(\mathbf{k},\mathbf{k}'):\mathbf{F}_{z\lessgtr z'}(z,z')}{k_z k^2_p} \right]e^{i\mathbf{k}_p\cdot(\mathbf{r}_p-\mathbf{r}'_p)},
\end{align}
where $\mu_{\mathrm{r}}(\omega)$ is the magnetic permeability in the layer with the emitting source, $k^2=k_z^2+k_p^2$, the operation $:$ corresponds to the Frobenius inner product, $\mathbf{F}_{z\lessgtr z'}(z,z')$ corresponds to a matrix that describes the waves traveling upward or downward in a given layer $n$, and
\begin{equation}
    \mathbf{C}(\mathbf{k},\mathbf{k}')= \begin{pmatrix}
    \mathbf{m}^*(\mathbf{k})\otimes \mathbf{m}(\mathbf{k'}) & \mathbf{n}^*(\mathbf{k})\otimes \mathbf{m}(\mathbf{k'}) \\ 
    \mathbf{m}^*(\mathbf{k})\otimes \mathbf{n}(\mathbf{k'}) & \mathbf{n}^*(\mathbf{k})\otimes \mathbf{n}(\mathbf{k'}) 
    \end{pmatrix},
\end{equation}

\noindent where $\mathbf{m}(\mathbf{r})= i\nabla_\mathbf{r}\times\hat{z}$ and $\mathbf{n}(\mathbf{r})=\frac{1}{k}\nabla_\mathbf{r}\times\nabla_\mathbf{r}\times\hat{z}$ are the dyadic operators that generate the s and p polarization vectors \cite{nano_optics_2012} and $\otimes$ is the outer product.

For only one interface below the radiation emitter, located at $z'$, we have that 
\begin{equation}
    \mathbf{F}_{z< z'}(z,z') = \mathbf{F}_{z> z'}(z,z') = e^{ik_z|z-z'|}\mathbf{I}+e^{ik_z(z+z')}\mathbf{R}.
\end{equation}
Note that if we introduce $\mathbf{F}(z,z')$ into equation \eqref{eq:general_gf} we recover the electric free-space dyadic Green's function $\mathbf{G}_0$, plus another extra term corresponding to the reflected dyadic Green's function $\mathbf{G}_R$,
\begin{equation}
    \mathbf{G}(\mathbf{r},\mathbf{r}',\omega) = \mathbf{G}_0(\mathbf{r},\mathbf{r}',\omega)+\mathbf{G}_R(\mathbf{r},\mathbf{r}',\omega).
\end{equation}

\section{Adding a Mu-Metal Layer}
\begin{figure}[!htp]
    \centering
    \includegraphics[width=0.9\linewidth]{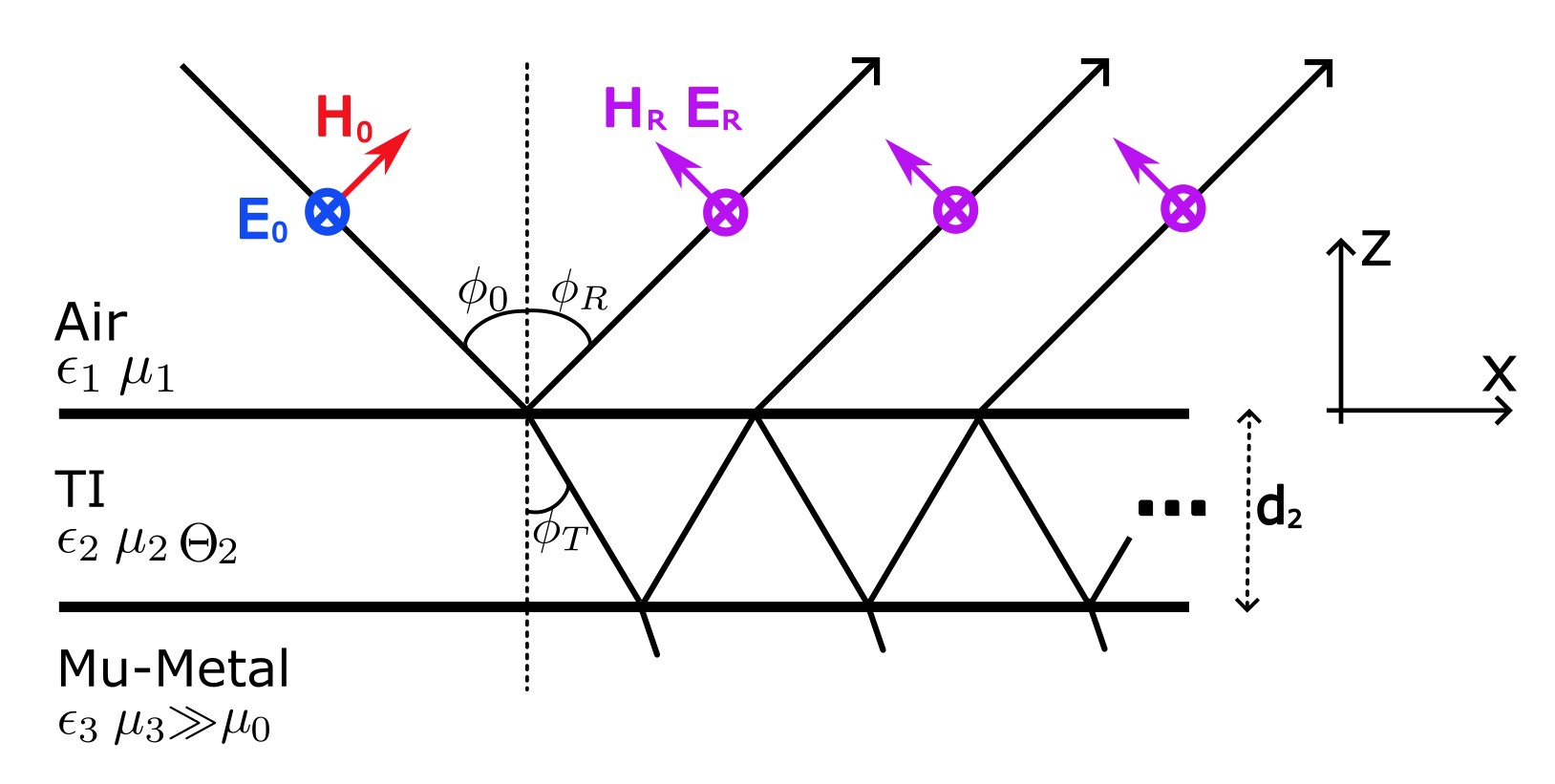}
    \caption{Sketch of a TE polarized ray impacting the three-layered Air-TI-Mu-Metal system, where the reflected wave has electric and magnetic field components in all directions due to the Electromagnetic coupling.}
    \label{fig:3_layer_sketch}
\end{figure}

As the intrinsic parameters of the TI are difficult to change without altering its properties, it is far more convenient to find a way to incorporate another material with a very high value for $\mu$ that enhances the mixed reflection coefficients similarly, as $\Delta \sim \mu_1\mu_2$. This addition increases the relative scale of these coefficients with respect to the usual coefficients. Thus, we propose to examine a three-layered system with a Mu-metal with $\mu \sim 10^{5}\mu_0$, as the one seen in figure \ref{fig:3_layer_sketch}. It remains to be seen where to set this layer: over or under the topological insulator. 

To determine the ordering of the layers, and how the reflection matrix will behave in each case, we study the reflection model for a 3-layered system \cite{weng_cho_waves}. By taking into account not only the first reflected wave but also the transmitted portion which is reflected on the second interface and transmitted once again through the first one, we can define a new reflection matrix, such that
\begin{equation}\label{eq:new_R_matrix}
    \mathbf{\tilde{R}}_{12}= \mathbf{R}_{12} + e^{2ik_{z2}(d_{12}-d_{23})}\mathbf{T}_{21}\cdot\mathbf{R}_{23}\cdot\mathbf{M}_{2123}^{-1}\cdot\mathbf{T}_{12}, 
\end{equation}
where
\begin{equation}\label{M-Matrix}
    \mathbf{M}_{2123} = \mathbf{I} - e^{2ik_{z2}(d_{12}-d_{23})}\mathbf{R}_{21}\cdot\mathbf{R}_{23}.
\end{equation}

If the emission source remains on the first layer of our configuration, it is only necessary to replace the reflection matrix in $\mathbf{F}_{z\lessgtr z'}(z,z')$ with \eqref{eq:new_R_matrix}. 

\subsection{3-Layers Fresnel Coefficients}
For the following numerical results, we used the optical properties of room-temperature TI TlBiSe$_2$\cite{PhysRevLett.105.136802, MitsasGrowth, Castro2020, Phutela2022}, as its relatively high impedance magnifies the topological effects previously discussed. These optical properties are $\epsilon\approx 4,\mu=1$. We also consider another TI with $\epsilon=1,\mu=2$, which can be achieved by increasing the magnetic permeability through doping\cite{Chen2019, Kim2019, Choi2011}. Magnetically doping TIs is a prominent technique for enhancing the anomalous magnetic effects displayed, despite the risk of losing the topological properties due to anisotropy in the Bulk\cite{Teng2019}. We use the latter optical properties in contrast to TlBiSe$_2$ to show relevant dependencies that arise with the TI's impedance. 

\begin{figure}[!htp]
\centering
\includegraphics[width=67mm]{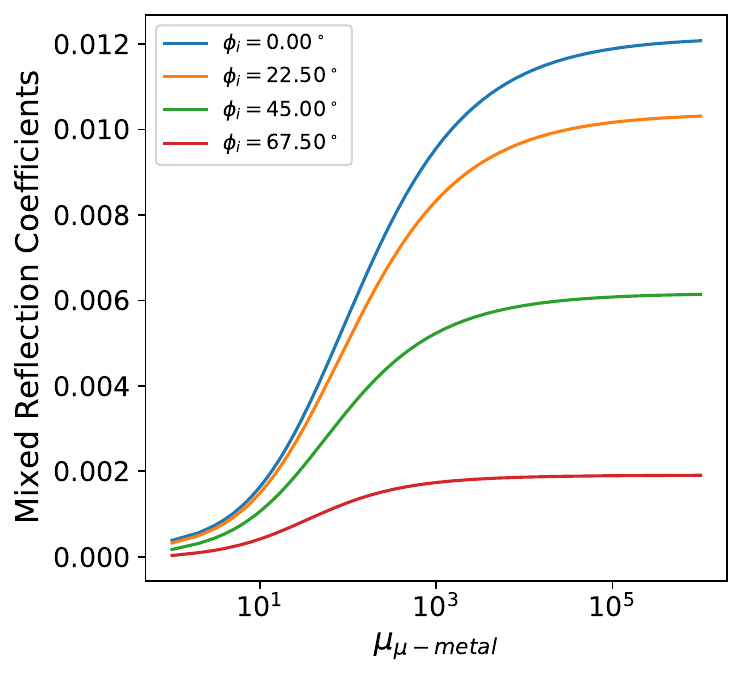}
\caption{Mixed reflective coefficients as a function of the $\mu$ of the mu-metal, for different incident angles, with $\mu_1=\epsilon_1=\epsilon_{\Theta}=1$, $\mu_{\Theta}=2$, $\lambda=600 \;\text{nm}$ and $d_{12}-d_{23}=100 \;\text{nm}$.}
\label{fig:new_parameters_mu-metal}
\end{figure}

\begin{figure}[!htp]
\begin{tabular}{cc}
  (a) & \\
  (b) & \includegraphics[width=67mm]{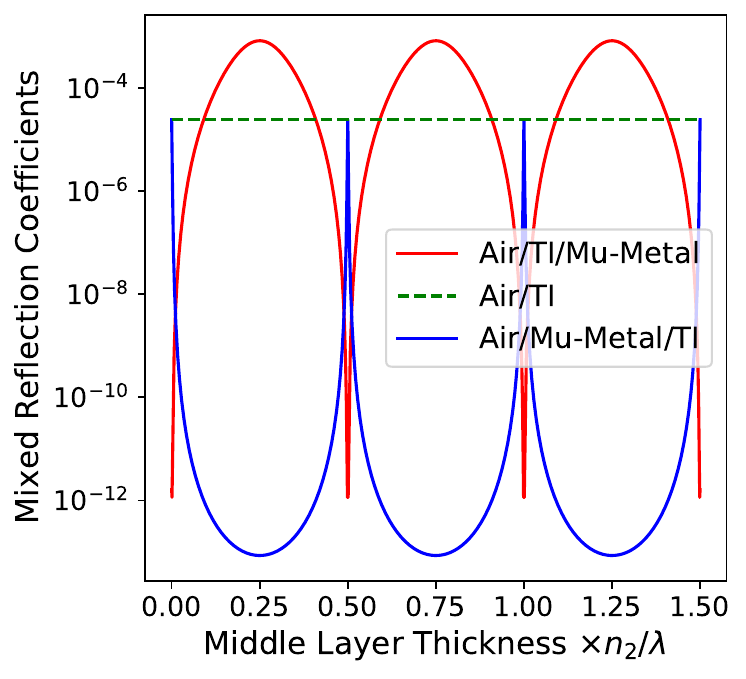} \\
   & \includegraphics[width=67mm]{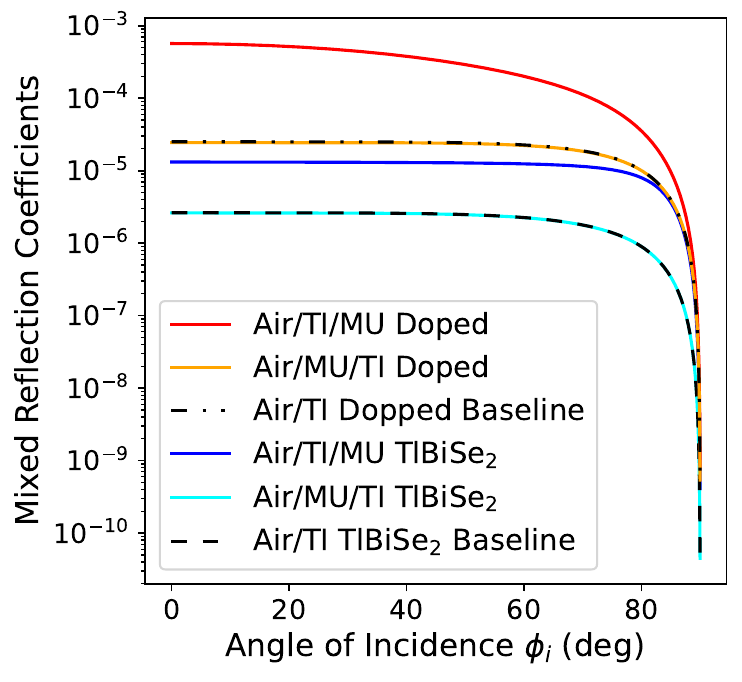} \\[6pt]
\end{tabular}
\caption{(a) Mixed reflective coefficients as a function of the TI layer's thickness, in units of the optical wavelength in the medium ($\lambda/n_2$), to visualize the same amount of peaks in both configurations. Here, the angle of incidence $\phi_i=0^\circ$, $\mu_{\Theta}=2$ and $\epsilon_{\Theta}=1$. (b) Mixed reflective coefficients as a function of the incident angle for the three different possible configurations for two different TIs, with $\mu_1=\epsilon_1=\epsilon_{\mu-\text{metal}}=1$, $\mu_{\mu-\text{metal}}=10^5$, $\lambda=600\; \text{nm}$ and middle layer thickness is strategically chosen to maximize them. The TI's parameters of TlBiSe$_2$ are $\mu_{\Theta}=1$ and $\epsilon_{\Theta}=4$, and for the doped TI are $\mu_{\Theta}=2$ and $\epsilon_{\Theta}=1$.}
\label{fig:reflexion_mu_metal}
\end{figure}

In figure \ref{fig:new_parameters_mu-metal} we can see how, as we increase the mu-metal $\mu$ value, the mixed reflective coefficients tend asymptotically to a maximum value, so after a certain $\mu$ the gain from increasing this value will diminish. As for the thickness of the topological insulator, due to the oscillatory nature of the exponential terms on equations (\ref{eq:new_R_matrix}) and (\ref{M-Matrix}), for a given angle of incidence over the surface, we can select an appropriate thickness on which the mixed reflection coefficients are at their maximum. In figure \ref{fig:reflexion_mu_metal}(a), we can see this behavior mirrored, and we can use these peaks to determine how the TI's parameters influence these increases.  

Here it is also clear that the Air/TI/Mu-Metal configuration's peaks deviate above from the baseline, while the Air/Mu-Metal/TI configuration peaks are at most at the base value. This last phenomenon can be attributed to Mu-Metal shielding\cite{yamazaki2005incremental, ter1991improvement, gao2023low, dubbers1986simple}, as starting to increase the thickness of the Mu-Metal in this configuration results in the total reflection of the light before it reaches the TI layer, causing any mixed reflection coefficients to nullify. This argument is sustained by the reflection and transmission matrices in table \ref{tab:coeff_limits}, for the Mu-Metal cases.

As seen in figure \ref{fig:reflexion_mu_metal}(b), for TI with lower impedance, we are able to enhance by approximately half an order of magnitude the mixed reflection coefficients introducing the Air/TI/Mu-Metal configuration, taking into account the optimal thickness for these parameters obtained numerically. The Air/Mu-Metal/TI configuration is found to peak at the base configuration of Air/TI. In figure \ref{fig:reflexion_mu_metal}(b), we also see that the increase is more significant as the TI's impedance is raised. Meanwhile, the maximum of the Air/Mu-Metal/TI does not improve with higher impedance. We predict mixed coefficients around $10^2$ times bigger than those without the Mu-Metal sub-layer. 

We did not consider absorption in our models. In this case, absorption would add an exponential decay to the reflection coefficients, which depends on the middle layer's thickness. Thus, selecting the thickness corresponding to the first peak in the coefficients is advised. 

We make the approximation $\mu_3\gg \mu_2,\mu_1,\epsilon_1,\epsilon_2,\epsilon_3$ in order to show the algebraic dependencies on the parameters. From \eqref{M-Matrix}, we obtain
\small
\begin{align}\label{eq:M_Matrix_explicit}
    &\mathbf{M}_{2123} = \nonumber \\
    &\left(\begin{matrix}
    1-\mathcal{F}(R_\textit{EE}^{21}R_\textit{EE}^{23} + R_\textit{EM}^{21}R_\textit{ME}^{23}) & -\mathcal{F}(R_\textit{EE}^{21}R_\textit{EM}^{23} + R_\textit{EM}^{21}R_\textit{MM}^{23})\\
    -\mathcal{F}(R_\textit{ME}^{21}R_\textit{EE}^{23} + R_\textit{MM}^{21}R_\textit{ME}^{23}) & 1-\mathcal{F}(R_\textit{ME}^{21}R_\textit{EM}^{23} + R_\textit{MM}^{21}R_\textit{MM}^{23})
    \end{matrix}\right),
\end{align}
\normalsize
where $\mathcal{F}=e^{2ik_{z2}(d_{12}-d_{23})}$ and, for example, $R_\textit{EE}^{21}$ is the $R_\textit{TE,TE}$ entry of the $\mathbf{R}_{21}$ reflection matrix. Now, using our approximation, we can reduce the $\mathbf{R}_{23}$ such that,
\begin{equation}
    \mathbf{R}_{23} \approx \left(\begin{matrix}
    1 & -2 Z_2 \alpha (\Theta_3-\Theta_2) \\ -2 Z_2 \alpha (\Theta_3-\Theta_2) & -1
    \end{matrix}\right),
\end{equation}

$Z_2=\sqrt{\mu_2/\epsilon_2}$ being the TI's impedance. Multiplying \eqref{eq:M_Matrix_explicit} with this approximation, we obtain the following matrix,
\begin{align}\label{eq:beta_lin}
    &\mathbf{R}_{23}\cdot\mathbf{M}_{2123}^{-1} = \nonumber\\
    &\frac{1}{\det(\mathbf{M}_{2123})}\left(\begin{matrix}
    1 & -\beta \\ -\beta & -1
    \end{matrix}\right) + \frac{(1+\beta^2)\mathcal{F}}{\det(\mathbf{M}_{2123})}\left(\begin{matrix}
    R_\textit{MM}^{21} & R_\textit{EM}^{21} \\ R_\textit{ME}^{21} & R_\textit{EE}^{21}
    \end{matrix}\right),
\end{align}
where $\beta = 2 Z_2 \alpha (\Theta_3-\Theta_2)/\pi$. It is worth remarking that new terms appear on the off-diagonal components of our matrix that are dependant on $\beta/\det(\mathbf{M}_{2123})$. Given that $\beta \sim 10^{-2}$ in order, and $\det(\mathbf{M}_{2123}) \sim 1$, these new terms should dominate over the mixed reflection coefficients that are of order $10^{-3}$. Given this, we roughly estimate the order of the new reflection coefficients with equations \eqref{eq:new_R_matrix} and \eqref{eq:real_coefficients}. We see that the maximum of the new reflection coefficients is of order $10^{-4}$, which is shown in figure \ref{fig:reflexion_mu_metal}.

Therefore, we can confirm that the increase in the mixed reflection coefficients, due to the addition of a  Mu-Metal layer, directly correlates to the off-diagonal terms in the first term of equation \eqref{eq:beta_lin}. The increase scales linearly with the TI's impedance, as long as $\mu_3\gg\mu_2$ is satisfied.

\subsection{3-Layers Green's Function/Electric Field}
In this section, we will display the results obtained for the Green's function over the TI. From equation \eqref{eq:E_and_G_with_J}, we can understand each column of $\mathbf{G}(\mathbf{r},\mathbf{r}',\omega)$ as a scaled representation of the directional real components of the electric field generated by an oscillating dipole on $\mathbf{r}'$, oriented along the $\hat{x}, \hat{y}$ and $\hat{z}$ directions, as for that case
\begin{equation}
    \mathbf{J}(\mathbf{r}')=-i\omega d(\omega)\delta(\mathbf{r}')\hat{e}_\mu,
\end{equation} 
with $\hat{e}_\mu$ the unitary dipole orientation and $d(\omega)$ the dipole strength. We obtain that the electric field is proportional by a factor to $\mathbf{G}(\mathbf{r},\mathbf{r}',\omega)\cdot\hat{e}_\mu$. 

For free-space and Air/Magnetodielectric configurations, for an $\hat{x}$ oriented dipole, the electric field $\hat{y}$ component in the $x-z$ plane is null, and the $\hat{x}$ and $\hat{z}$ components are not. This can be seen in figures \ref{fig:G0_plot} and \ref{fig:Gmd_plot} in appendix \ref{appendix:G_graphs} and is crucial to characterize the difference between magnetodielectric materials and topological insulators, as this will not continue to be true for topologically complex materials. 

\begin{figure}
    \centering
    \includegraphics[width=1.03\linewidth]{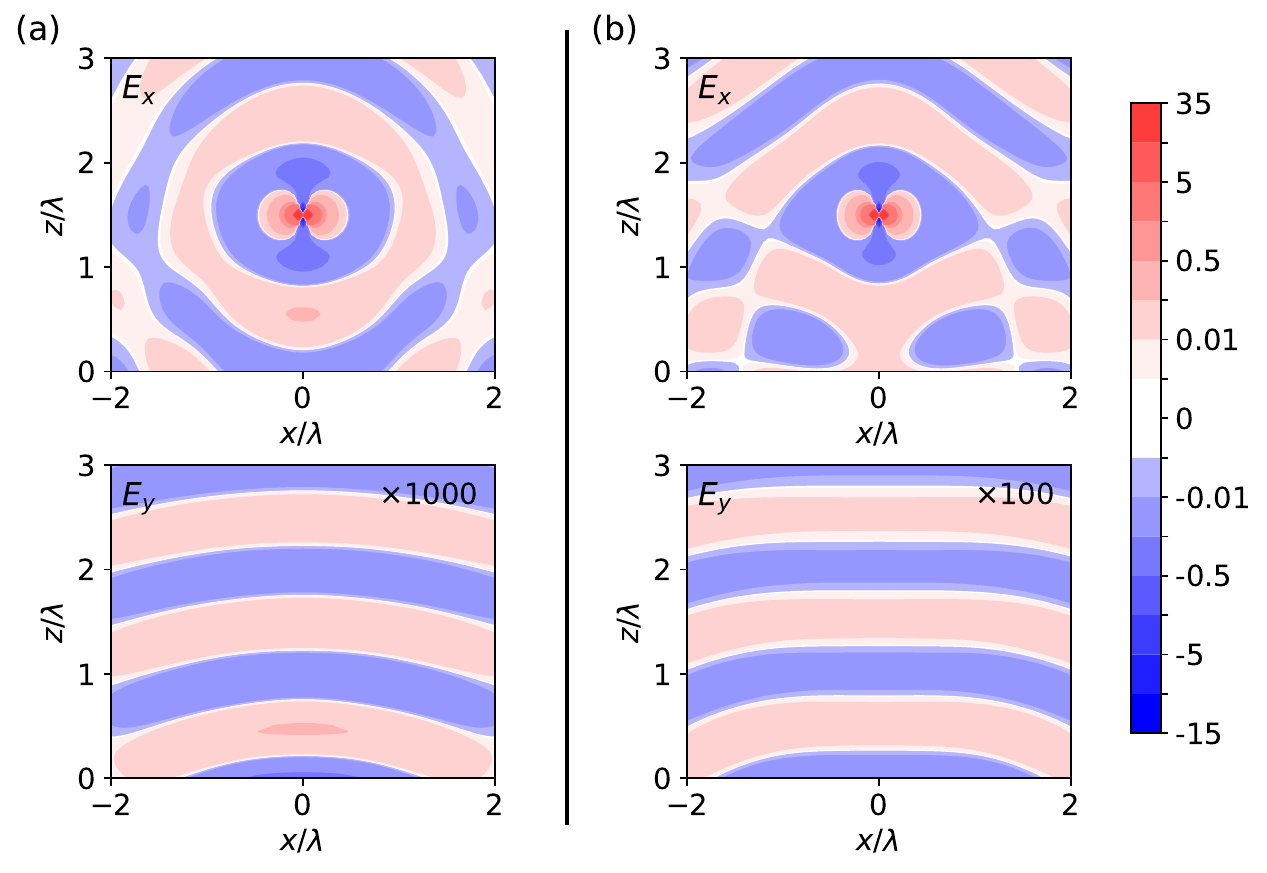}
    \caption{Electric field components $\mathbf{E}_{i}\times\lambda$, graphed in the $y=0$ plane, as a function of $z/\lambda$ and $x/\lambda$ for (a) an Air/TI configuration, and (b) an Air/TI/Mu-Metal configuration. $E_y$ is solely due to the topological response of the configuration and had to be enhanced to be observed with the non-topological $E_x$. Note that the necessary enhancement factor is reduced by an order of magnitude when adding a Mu-Metal to the system. The corresponding parameters, other components for each figure, and the components for the free-space and Air/Magnetodielectric configurations can be found in appendix \ref{appendix:G_graphs}.}
    \label{fig:E_for_configurations}
\end{figure}

Now, analyzing the 2-layer Air/TI system configuration, as we can see in figure \ref{fig:E_for_configurations} (a), the mixed reflected coefficients caused the y-combined matrix components to appear, which can be physically seen by measuring, for example, the $y$-oriented component of the electric field of an $x$ or $z$-oriented dipole near the surface. Given the relatively small value of the mixed reflection coefficients, the mixed components of the matrix have been scaled by a $\times1000$ factor to be able to appreciate them in the same magnitude as the non-mixed entries. 

As previously discussed, the mixed Fresnel coefficients can be enhanced by implementing an Air/TI/Mu-metal 3-layered configuration. In figure \ref{fig:E_for_configurations} (b), we can see how the mixed entries of the electric field are now scaled by a $\times100$ factor, which means we have reduced by an order of magnitude the threshold necessary to be able to measure this effect via the electric field in an experimental environment. We can assure that this enhancement of the effect can be greater for different values of $\mu$ for the topological insulator layer, for which we can determine the optimal thickness of the TI, as discussed in the previous section.   

\section{New Component on Poynting Vector}
During previous sections, we have studied how the electric and magnetic fields, upon reflection or transmission on a TI, present components not only in their original polarization but also in the complementary polarization. In this section, we will study if this is also the case for the Poynting vector. If there is a substantial deviation, this might be the basis for a new way to characterize and measure this effect.


So, we are interested in the time-averaged Poynting vector, given by 
\begin{equation}\label{eq:poynting_vec}
    \mathbf{S} = \frac{1}{2}\text{Re}\{\mathbf{E}\times\mathbf{H}^*\}.
\end{equation}

\begin{figure*}[!ht]
\centering
\includegraphics[width=\linewidth]{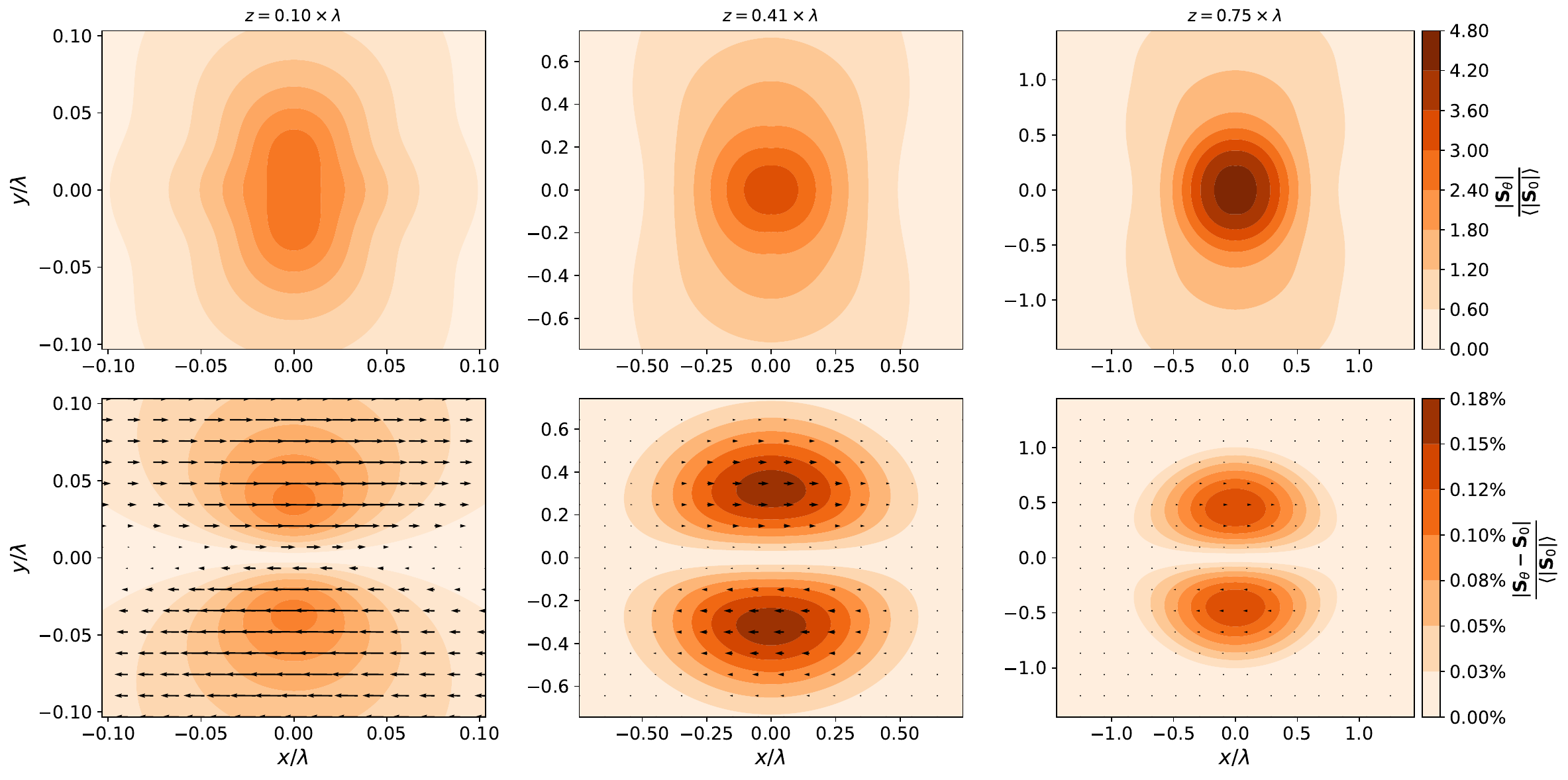}
\caption{Poynting vector (first row) and the difference between the Poynting vector (second row) generated by an $\hat{x}$ oriented dipole located at $z'=0.05\times\lambda$ over a TI's surface $\mathbf{S_\theta}$, and over the equivalent magnetodielectric surface $\mathbf{S_0}$, normalized by the average of the non-topological Poynting vector over the plane. Simulating an 0.9 numerical aperture (NA) at $z=0.1\times\lambda$, $0.35\times\lambda$, and $0.75\times\lambda$ over the TI's surface. All plots consider the parameters $\mu_1=\epsilon_1=\mu_\Theta=1$, $\epsilon_\Theta=4$, $\lambda=600 \;\text{nm}$, and $\Theta_2=\pi$ for the TI.}
\label{fig:real_deviation}
\end{figure*}

Upon taking the ansatz for a single TE polarized planar wave and calculating the corresponding Poyinting vector for the top layer of an Air-TI configuration, we obtain the following solution,
\small \begin{equation}\label{eq:s_te}
    \mathbf{S} = \frac{E_0^2}{2\mu_0\mu_1\omega}\left(\begin{matrix}
    k_p\left[1+R_\textit{TE,TE}^2 + 2\cos{(2k_z)}R_\textit{TE,TE}- R_\textit{TM,TE}^2\right]\\
    \frac{k_{z1} k_p}{k_1}R_\textit{TE,TE}\;R_\textit{TM,TE}\\
    k_{z1}\left[1+R_\textit{TE,TE}^2 + R_\textit{TM,TE}^2\right]
    \end{matrix}\right).
\end{equation}
\normalsize
The important feature of equation \eqref{eq:s_te} is its $\hat{y}$ component, which is nullified for regular magnetodielectric media. This new component suggests a relatively small transversal deviation of the Poynting vector.  

For a dipole over a TI's surface, we can expand the analysis for the Green's function to also obtain the magnetic field, given by\cite{nano_optics_2012} 
\begin{equation}
    \mathbf{H}(\mathbf{r},\omega)=\int d^3r'\left[\nabla\times \mathbf{G}(\mathbf{r},\mathbf{r}',\omega)\right]\cdot\mathbf{J}(\mathbf{r}').
\end{equation}
Expanding on the result for the electric Green's function\cite{TSB-TI}, computing its rotor, and using the corresponding Hankel transform\cite{Gaskill1978-yg}, we are able to add the reflected part of the magnetic Green's function for a TI half-plane space. Here are the results for an $\hat{x}$ oriented dipole:

\begin{widetext}
\small
\begin{align}
    \label{eq:M_xx} M^{\,xx}_R&(\mathbf{r},\mathbf{r}',\omega) = \frac{i\mu(\omega)}{8\pi} \int_0^{k_1}dk_p k_p \;e^{ik_z(|z|+|z'|)}
     \left[J_0(k_pR_p)\left(\frac{k_1}{k_{z}}+\frac{k_{z}}{k_1}\right) - J_2(k_pR_p)\frac{k_p^2}{k_1\,k_z}\right] R_{\text{TE,TM}},\\
    \label{eq:M_xy} M^{\,yx}_R&(\mathbf{r},\mathbf{r}',\omega) = \frac{i\mu(\omega)}{8\pi} \int_0^{k_1}dk_pk_p\;e^{ik_z(|z|+|z'|)}
    \Big[(J_0(k_pR_p)+J_2(k_pR_p))(R_{\text{TE,TE}}+R_{\text{TM,TM}}) +
     J_2(k_pR_p)\frac{2k_pk_z^2}{k_1^2}R_{\text{TM,TM}}\Big],\\
    \label{eq:M_xz} M^{\,zx}_R&(\mathbf{r},\mathbf{r}',\omega) = \frac{\mu(\omega)}{4\pi} \int_0^{k_1}dk_pe^{ik_z(|z|+|z'|)}J_1(k_pR_p)\;\frac{k_p^2}{k_1^2} R_{\text{TE,TM}},
\end{align}
\end{widetext}

\normalsize
\noindent where $k_z=\sqrt{k_1^2-k_p^2}$, and $J_0$, $J_1$ and $J_2$ are first kind Bessel functions. This can also be extended to the rest of the dipole orientations. Substituting these results and the electrical field's Green's function into equation \eqref{eq:s_te} and computing the difference between this vector field for TI and its non-topological counterpart $\mathbf{S}_\Theta - \mathbf{S}_0$, we obtain the graph on figure \ref{fig:real_deviation}. Here we see a clear new transverse component on the $\hat{y}$ direction, due to the topological effect. 

\begin{figure}[!htp]
    \centering
    \includegraphics[width=0.8\linewidth]{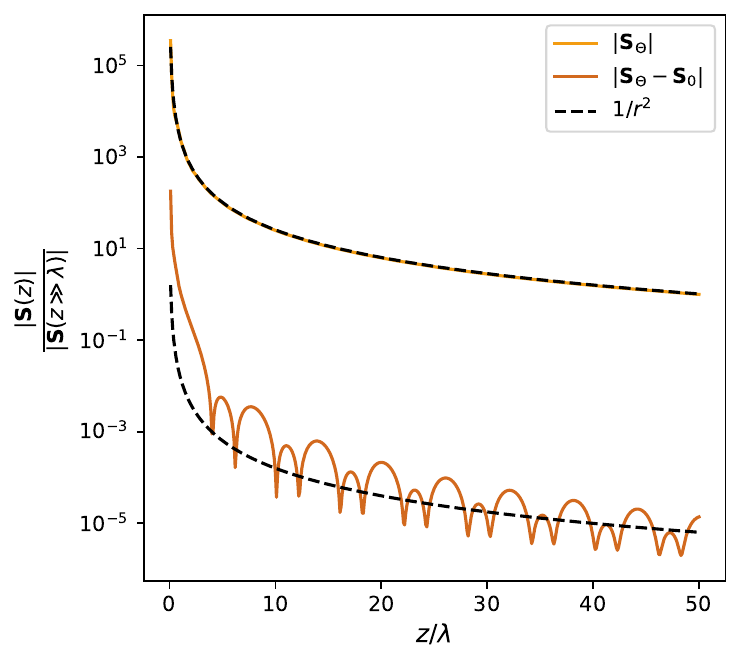}
    \caption{Topological Poynting vector ($\mathbf{S}_\Theta$) and the topological Poynting vector's deviation ($\mathbf{S}_\Theta-\mathbf{S}_0$) normalized by $\mathbf{S}_\Theta$ on the far-field. Note that both have components that decay as $1/r^2$. The oscillatory behavior on the deviation is due to the Bessel functions dependency of $\mathbf{S}_\Theta$.}
    \label{fig:far_field_poynting}
\end{figure}

We also notice that the relative difference tends to decrease the further we go from the emission source. Looking at equations \eqref{eq:M_xx}, \eqref{eq:M_xy} and \eqref{eq:M_xz}, specifically their dependence on the Bessel functions $J_n(k_p\cdot R_p)$ and the exponential, we expect that the results for the Poynting vector should scale by $\sim 1/r^2$, as evidenced in figure \ref{fig:far_field_poynting}. This indicates that the relative difference is indeed a far-field effect, as the integral over the solid angle should be constant, as in the case of the non-topological Poynting vector.

Moreover, for a room temperature TI with a lower impedance, we should expect this effect to be lower, although we are able to find the optimal configuration of the dipole-TI system to find the maximum possible percentual deviation. Figure \ref{fig:real_deviation} shows a maximum transversal deviation of $0.18\%$ over the expected Poynting vector of the equivalent magnetodielectric.

\section{Conclusions}
We considered the inclusion of a  Mu-Metal layer in the system Air/Topological Insulator to enhance the optical effects of the mixed reflection Fresnel coefficients. We found that it is most beneficial to situate it under the Topological Insulator layer and that the increase is more substantial for TIs with higher impedance. We also discussed how to find the optimal thickness for the TI middle layer, given the angle of incidence and the TI's properties. In our simulations, we predict an increase of $\sim \times 100$ for a TI with impedance $Z=\sqrt{2}$.  

Finally, we extended the electric Green's function analysis to obtain the magnetic Green's function and the Poynting vector field for a dipole over a TI. We found that the Poynting vector presents a new transverse component due to the new mixed Fresnel coefficients, which scale like $1/r^2$, where $r$ is the distance to the source, indicating this effect will be present on the far field. Moreover, for a room-temperature TI, we predict a maximum percentual deviation of $0.18\%$, which indicates that this effect is measurable given the right experimental setup. 

Since the result showed that the new magnetic components that appear in TIs are enhanced, we can more confidently expect that magnetic probing, using single emitters, might prove a reliable way to characterize TIs. Moreover, the enhancement shown by adding Mu-Metal materials may be aided with additional meta-materials in a variety of geometrical forms being introduced into the layered configuration to enhance these effects further. 

\section{ACKNOWLEDGEMENTS}
The authors acknowledge the support from the Asian Office of Aerospace Research and Development (AOARD) FA2386-21-1-4125, Fondecyt Regular No 1221512, and Anillo ACT192023 for this work. 

\newpage
\bibliographystyle{apsrev4-1}
\bibliography{bib}

\begin{thebibliography}{51}%
\makeatletter
\providecommand \@ifxundefined [1]{%
 \@ifx{#1\undefined}
}%
\providecommand \@ifnum [1]{%
 \ifnum #1\expandafter \@firstoftwo
 \else \expandafter \@secondoftwo
 \fi
}%
\providecommand \@ifx [1]{%
 \ifx #1\expandafter \@firstoftwo
 \else \expandafter \@secondoftwo
 \fi
}%
\providecommand \natexlab [1]{#1}%
\providecommand \enquote  [1]{``#1''}%
\providecommand \bibnamefont  [1]{#1}%
\providecommand \bibfnamefont [1]{#1}%
\providecommand \citenamefont [1]{#1}%
\providecommand \href@noop [0]{\@secondoftwo}%
\providecommand \href [0]{\begingroup \@sanitize@url \@href}%
\providecommand \@href[1]{\@@startlink{#1}\@@href}%
\providecommand \@@href[1]{\endgroup#1\@@endlink}%
\providecommand \@sanitize@url [0]{\catcode `\\12\catcode `\$12\catcode
  `\&12\catcode `\#12\catcode `\^12\catcode `\_12\catcode `\%12\relax}%
\providecommand \@@startlink[1]{}%
\providecommand \@@endlink[0]{}%
\providecommand \url  [0]{\begingroup\@sanitize@url \@url }%
\providecommand \@url [1]{\endgroup\@href {#1}{\urlprefix }}%
\providecommand \urlprefix  [0]{URL }%
\providecommand \Eprint [0]{\href }%
\providecommand \doibase [0]{http://dx.doi.org/}%
\providecommand \selectlanguage [0]{\@gobble}%
\providecommand \bibinfo  [0]{\@secondoftwo}%
\providecommand \bibfield  [0]{\@secondoftwo}%
\providecommand \translation [1]{[#1]}%
\providecommand \BibitemOpen [0]{}%
\providecommand \bibitemStop [0]{}%
\providecommand \bibitemNoStop [0]{.\EOS\space}%
\providecommand \EOS [0]{\spacefactor3000\relax}%
\providecommand \BibitemShut  [1]{\csname bibitem#1\endcsname}%
\let\auto@bib@innerbib\@empty
\bibitem [{\citenamefont {Qi}\ and\ \citenamefont {Zhang}(2011)}]{Qi_Zhang}%
  \BibitemOpen
  \bibfield  {author} {\bibinfo {author} {\bibfnamefont {X.-L.}\ \bibnamefont
  {Qi}}\ and\ \bibinfo {author} {\bibfnamefont {S.-C.}\ \bibnamefont {Zhang}},\
  }\href {\doibase 10.1103/RevModPhys.83.1057} {\bibfield  {journal} {\bibinfo
  {journal} {Rev. Mod. Phys.}\ }\textbf {\bibinfo {volume} {83}},\ \bibinfo
  {pages} {1057} (\bibinfo {year} {2011})}\BibitemShut {NoStop}%
\bibitem [{\citenamefont {Hasan}\ and\ \citenamefont
  {Moore}(2011)}]{Hasan_Moore2011}%
  \BibitemOpen
  \bibfield  {author} {\bibinfo {author} {\bibfnamefont {M.~Z.}\ \bibnamefont
  {Hasan}}\ and\ \bibinfo {author} {\bibfnamefont {J.~E.}\ \bibnamefont
  {Moore}},\ }\href {\doibase 10.1146/annurev-conmatphys-062910-140432}
  {\bibfield  {journal} {\bibinfo  {journal} {Annual Review of Condensed Matter
  Physics}\ }\textbf {\bibinfo {volume} {2}},\ \bibinfo {pages} {55} (\bibinfo
  {year} {2011})}\BibitemShut {NoStop}%
\bibitem [{\citenamefont {Ando}(2013)}]{Yoichi2013}%
  \BibitemOpen
  \bibfield  {author} {\bibinfo {author} {\bibfnamefont {Y.}~\bibnamefont
  {Ando}},\ }\href {\doibase 10.7566/JPSJ.82.102001} {\bibfield  {journal}
  {\bibinfo  {journal} {Journal of the Physical Society of Japan}\ }\textbf
  {\bibinfo {volume} {82}},\ \bibinfo {pages} {102001} (\bibinfo {year}
  {2013})}\BibitemShut {NoStop}%
\bibitem [{\citenamefont {Qi}\ \emph {et~al.}(2009)\citenamefont {Qi},
  \citenamefont {Li}, \citenamefont {Zang},\ and\ \citenamefont
  {Zhang}}]{point_charge}%
  \BibitemOpen
  \bibfield  {author} {\bibinfo {author} {\bibfnamefont {X.-L.}\ \bibnamefont
  {Qi}}, \bibinfo {author} {\bibfnamefont {R.}~\bibnamefont {Li}}, \bibinfo
  {author} {\bibfnamefont {J.}~\bibnamefont {Zang}}, \ and\ \bibinfo {author}
  {\bibfnamefont {S.-C.}\ \bibnamefont {Zhang}},\ }\href {\doibase
  10.1126/science.1167747} {\bibfield  {journal} {\bibinfo  {journal}
  {Science}\ }\textbf {\bibinfo {volume} {323}},\ \bibinfo {pages} {1184}
  (\bibinfo {year} {2009})}\BibitemShut {NoStop}%
\bibitem [{\citenamefont {Essin}\ \emph {et~al.}(2009)\citenamefont {Essin},
  \citenamefont {Moore},\ and\ \citenamefont {Vanderbilt}}]{Essin_2009}%
  \BibitemOpen
  \bibfield  {author} {\bibinfo {author} {\bibfnamefont {A.~M.}\ \bibnamefont
  {Essin}}, \bibinfo {author} {\bibfnamefont {J.~E.}\ \bibnamefont {Moore}}, \
  and\ \bibinfo {author} {\bibfnamefont {D.}~\bibnamefont {Vanderbilt}},\
  }\href {\doibase 10.1103/physrevlett.102.146805} {\bibfield  {journal}
  {\bibinfo  {journal} {Physical Review Letters}\ }\textbf {\bibinfo {volume}
  {102}} (\bibinfo {year} {2009}),\ 10.1103/physrevlett.102.146805}\BibitemShut
  {NoStop}%
\bibitem [{\citenamefont {Mart\'{\i}n-Ruiz}\ \emph {et~al.}(2019)\citenamefont
  {Mart\'{\i}n-Ruiz}, \citenamefont {Rodr\'{\i}guez-Tzompantzi}, \citenamefont
  {Maze},\ and\ \citenamefont {Urrutia}}]{jero+martin}%
  \BibitemOpen
  \bibfield  {author} {\bibinfo {author} {\bibfnamefont {A.}~\bibnamefont
  {Mart\'{\i}n-Ruiz}}, \bibinfo {author} {\bibfnamefont {O.}~\bibnamefont
  {Rodr\'{\i}guez-Tzompantzi}}, \bibinfo {author} {\bibfnamefont {J.~R.}\
  \bibnamefont {Maze}}, \ and\ \bibinfo {author} {\bibfnamefont {L.~F.}\
  \bibnamefont {Urrutia}},\ }\href {\doibase 10.1103/PhysRevA.100.042124}
  {\bibfield  {journal} {\bibinfo  {journal} {Phys. Rev. A}\ }\textbf {\bibinfo
  {volume} {100}},\ \bibinfo {pages} {042124} (\bibinfo {year}
  {2019})}\BibitemShut {NoStop}%
\bibitem [{\citenamefont {Kitagawa}\ \emph {et~al.}(2010)\citenamefont
  {Kitagawa}, \citenamefont {Hiraoka}, \citenamefont {Honda}, \citenamefont
  {Ishikura}, \citenamefont {Nakamura},\ and\ \citenamefont
  {Kimura}}]{Kitagawa2010}%
  \BibitemOpen
  \bibfield  {author} {\bibinfo {author} {\bibfnamefont {Y.}~\bibnamefont
  {Kitagawa}}, \bibinfo {author} {\bibfnamefont {Y.}~\bibnamefont {Hiraoka}},
  \bibinfo {author} {\bibfnamefont {T.}~\bibnamefont {Honda}}, \bibinfo
  {author} {\bibfnamefont {T.}~\bibnamefont {Ishikura}}, \bibinfo {author}
  {\bibfnamefont {H.}~\bibnamefont {Nakamura}}, \ and\ \bibinfo {author}
  {\bibfnamefont {T.}~\bibnamefont {Kimura}},\ }\href {\doibase
  10.1038/nmat2826} {\bibfield  {journal} {\bibinfo  {journal} {Nature
  Materials}\ }\textbf {\bibinfo {volume} {9}},\ \bibinfo {pages} {797}
  (\bibinfo {year} {2010})}\BibitemShut {NoStop}%
\bibitem [{\citenamefont {Velev}\ \emph {et~al.}(2011)\citenamefont {Velev},
  \citenamefont {Jaswal},\ and\ \citenamefont {Tsymbal}}]{multiferroic}%
  \BibitemOpen
  \bibfield  {author} {\bibinfo {author} {\bibfnamefont {J.~P.}\ \bibnamefont
  {Velev}}, \bibinfo {author} {\bibfnamefont {S.~S.}\ \bibnamefont {Jaswal}}, \
  and\ \bibinfo {author} {\bibfnamefont {E.~Y.}\ \bibnamefont {Tsymbal}},\
  }\href {http://www.jstor.org/stable/23035864} {\bibfield  {journal} {\bibinfo
   {journal} {Philosophical Transactions: Mathematical, Physical and
  Engineering Sciences}\ }\textbf {\bibinfo {volume} {369}},\ \bibinfo {pages}
  {3069} (\bibinfo {year} {2011})}\BibitemShut {NoStop}%
\bibitem [{\citenamefont {Moore}(2010)}]{Moore2010}%
  \BibitemOpen
  \bibfield  {author} {\bibinfo {author} {\bibfnamefont {J.~E.}\ \bibnamefont
  {Moore}},\ }\href {\doibase 10.1038/nature08916} {\bibfield  {journal}
  {\bibinfo  {journal} {Nature}\ }\textbf {\bibinfo {volume} {464}},\ \bibinfo
  {pages} {194} (\bibinfo {year} {2010})}\BibitemShut {NoStop}%
\bibitem [{\citenamefont {Legg}\ \emph {et~al.}(2022)\citenamefont {Legg},
  \citenamefont {R\"{o}{\ss}ler}, \citenamefont {M\"{u}nning}, \citenamefont
  {Fan}, \citenamefont {Breunig}, \citenamefont {Bliesener}, \citenamefont
  {Lippertz}, \citenamefont {Uday}, \citenamefont {Taskin}, \citenamefont
  {Loss}, \citenamefont {Klinovaja},\ and\ \citenamefont {Ando}}]{Legg2022}%
  \BibitemOpen
  \bibfield  {author} {\bibinfo {author} {\bibfnamefont {H.~F.}\ \bibnamefont
  {Legg}}, \bibinfo {author} {\bibfnamefont {M.}~\bibnamefont
  {R\"{o}{\ss}ler}}, \bibinfo {author} {\bibfnamefont {F.}~\bibnamefont
  {M\"{u}nning}}, \bibinfo {author} {\bibfnamefont {D.}~\bibnamefont {Fan}},
  \bibinfo {author} {\bibfnamefont {O.}~\bibnamefont {Breunig}}, \bibinfo
  {author} {\bibfnamefont {A.}~\bibnamefont {Bliesener}}, \bibinfo {author}
  {\bibfnamefont {G.}~\bibnamefont {Lippertz}}, \bibinfo {author}
  {\bibfnamefont {A.}~\bibnamefont {Uday}}, \bibinfo {author} {\bibfnamefont
  {A.~A.}\ \bibnamefont {Taskin}}, \bibinfo {author} {\bibfnamefont
  {D.}~\bibnamefont {Loss}}, \bibinfo {author} {\bibfnamefont {J.}~\bibnamefont
  {Klinovaja}}, \ and\ \bibinfo {author} {\bibfnamefont {Y.}~\bibnamefont
  {Ando}},\ }\href {\doibase 10.1038/s41565-022-01124-1} {\bibfield  {journal}
  {\bibinfo  {journal} {Nature Nanotechnology}\ }\textbf {\bibinfo {volume}
  {17}},\ \bibinfo {pages} {696} (\bibinfo {year} {2022})}\BibitemShut
  {NoStop}%
\bibitem [{\citenamefont {Fischer}\ \emph {et~al.}(2022)\citenamefont
  {Fischer}, \citenamefont {Pic\'o-Cort\'es}, \citenamefont {Himmler},
  \citenamefont {Platero}, \citenamefont {Grifoni}, \citenamefont {Kozlov},
  \citenamefont {Mikhailov}, \citenamefont {Dvoretsky}, \citenamefont
  {Strunk},\ and\ \citenamefont {Weiss}}]{PhysRevResearch.4.013087}%
  \BibitemOpen
  \bibfield  {author} {\bibinfo {author} {\bibfnamefont {R.}~\bibnamefont
  {Fischer}}, \bibinfo {author} {\bibfnamefont {J.}~\bibnamefont
  {Pic\'o-Cort\'es}}, \bibinfo {author} {\bibfnamefont {W.}~\bibnamefont
  {Himmler}}, \bibinfo {author} {\bibfnamefont {G.}~\bibnamefont {Platero}},
  \bibinfo {author} {\bibfnamefont {M.}~\bibnamefont {Grifoni}}, \bibinfo
  {author} {\bibfnamefont {D.~A.}\ \bibnamefont {Kozlov}}, \bibinfo {author}
  {\bibfnamefont {N.~N.}\ \bibnamefont {Mikhailov}}, \bibinfo {author}
  {\bibfnamefont {S.~A.}\ \bibnamefont {Dvoretsky}}, \bibinfo {author}
  {\bibfnamefont {C.}~\bibnamefont {Strunk}}, \ and\ \bibinfo {author}
  {\bibfnamefont {D.}~\bibnamefont {Weiss}},\ }\href {\doibase
  10.1103/PhysRevResearch.4.013087} {\bibfield  {journal} {\bibinfo  {journal}
  {Phys. Rev. Res.}\ }\textbf {\bibinfo {volume} {4}},\ \bibinfo {pages}
  {013087} (\bibinfo {year} {2022})}\BibitemShut {NoStop}%
\bibitem [{\citenamefont {M\"{u}nning}\ \emph {et~al.}(2021)\citenamefont
  {M\"{u}nning}, \citenamefont {Breunig}, \citenamefont {Legg}, \citenamefont
  {Roitsch}, \citenamefont {Fan}, \citenamefont {R\"{o}{\ss}ler}, \citenamefont
  {Rosch},\ and\ \citenamefont {Ando}}]{Mnning2021}%
  \BibitemOpen
  \bibfield  {author} {\bibinfo {author} {\bibfnamefont {F.}~\bibnamefont
  {M\"{u}nning}}, \bibinfo {author} {\bibfnamefont {O.}~\bibnamefont
  {Breunig}}, \bibinfo {author} {\bibfnamefont {H.~F.}\ \bibnamefont {Legg}},
  \bibinfo {author} {\bibfnamefont {S.}~\bibnamefont {Roitsch}}, \bibinfo
  {author} {\bibfnamefont {D.}~\bibnamefont {Fan}}, \bibinfo {author}
  {\bibfnamefont {M.}~\bibnamefont {R\"{o}{\ss}ler}}, \bibinfo {author}
  {\bibfnamefont {A.}~\bibnamefont {Rosch}}, \ and\ \bibinfo {author}
  {\bibfnamefont {Y.}~\bibnamefont {Ando}},\ }\href {\doibase
  10.1038/s41467-021-21230-3} {\bibfield  {journal} {\bibinfo  {journal}
  {Nature Communications}\ }\textbf {\bibinfo {volume} {12}} (\bibinfo {year}
  {2021}),\ 10.1038/s41467-021-21230-3}\BibitemShut {NoStop}%
\bibitem [{\citenamefont {Wu}\ \emph {et~al.}(2021)\citenamefont {Wu},
  \citenamefont {Chen}, \citenamefont {Zhang}, \citenamefont {He},
  \citenamefont {Nance}, \citenamefont {Guo}, \citenamefont {Sasaki},
  \citenamefont {Shirokura}, \citenamefont {Hai}, \citenamefont {Fang},
  \citenamefont {Razavi}, \citenamefont {Wong}, \citenamefont {Wen},
  \citenamefont {Ma}, \citenamefont {Yu}, \citenamefont {Carman}, \citenamefont
  {Han}, \citenamefont {Zhang},\ and\ \citenamefont {Wang}}]{Wu2021}%
  \BibitemOpen
  \bibfield  {author} {\bibinfo {author} {\bibfnamefont {H.}~\bibnamefont
  {Wu}}, \bibinfo {author} {\bibfnamefont {A.}~\bibnamefont {Chen}}, \bibinfo
  {author} {\bibfnamefont {P.}~\bibnamefont {Zhang}}, \bibinfo {author}
  {\bibfnamefont {H.}~\bibnamefont {He}}, \bibinfo {author} {\bibfnamefont
  {J.}~\bibnamefont {Nance}}, \bibinfo {author} {\bibfnamefont
  {C.}~\bibnamefont {Guo}}, \bibinfo {author} {\bibfnamefont {J.}~\bibnamefont
  {Sasaki}}, \bibinfo {author} {\bibfnamefont {T.}~\bibnamefont {Shirokura}},
  \bibinfo {author} {\bibfnamefont {P.~N.}\ \bibnamefont {Hai}}, \bibinfo
  {author} {\bibfnamefont {B.}~\bibnamefont {Fang}}, \bibinfo {author}
  {\bibfnamefont {S.~A.}\ \bibnamefont {Razavi}}, \bibinfo {author}
  {\bibfnamefont {K.}~\bibnamefont {Wong}}, \bibinfo {author} {\bibfnamefont
  {Y.}~\bibnamefont {Wen}}, \bibinfo {author} {\bibfnamefont {Y.}~\bibnamefont
  {Ma}}, \bibinfo {author} {\bibfnamefont {G.}~\bibnamefont {Yu}}, \bibinfo
  {author} {\bibfnamefont {G.~P.}\ \bibnamefont {Carman}}, \bibinfo {author}
  {\bibfnamefont {X.}~\bibnamefont {Han}}, \bibinfo {author} {\bibfnamefont
  {X.}~\bibnamefont {Zhang}}, \ and\ \bibinfo {author} {\bibfnamefont {K.~L.}\
  \bibnamefont {Wang}},\ }\href {\doibase 10.1038/s41467-021-26478-3}
  {\bibfield  {journal} {\bibinfo  {journal} {Nature Communications}\ }\textbf
  {\bibinfo {volume} {12}} (\bibinfo {year} {2021}),\
  10.1038/s41467-021-26478-3}\BibitemShut {NoStop}%
\bibitem [{\citenamefont {Castro-Enr{\'{\i}}quez}\ \emph
  {et~al.}(2022)\citenamefont {Castro-Enr{\'{\i}}quez}, \citenamefont
  {Mart{\'{\i}}n-Ruiz},\ and\ \citenamefont {Cambiaso}}]{CastroEnrquez2022}%
  \BibitemOpen
  \bibfield  {author} {\bibinfo {author} {\bibfnamefont {L.~A.}\ \bibnamefont
  {Castro-Enr{\'{\i}}quez}}, \bibinfo {author} {\bibfnamefont {A.}~\bibnamefont
  {Mart{\'{\i}}n-Ruiz}}, \ and\ \bibinfo {author} {\bibfnamefont
  {M.}~\bibnamefont {Cambiaso}},\ }\href {\doibase 10.1038/s41598-022-24939-3}
  {\bibfield  {journal} {\bibinfo  {journal} {Scientific Reports}\ }\textbf
  {\bibinfo {volume} {12}} (\bibinfo {year} {2022}),\
  10.1038/s41598-022-24939-3}\BibitemShut {NoStop}%
\bibitem [{\citenamefont {Lin}\ \emph {et~al.}(2015)\citenamefont {Lin},
  \citenamefont {Lin}, \citenamefont {Chi}, \citenamefont {Wu}, \citenamefont
  {Cheng}, \citenamefont {Tseng}, \citenamefont {He}, \citenamefont {Wu},
  \citenamefont {Lee},\ and\ \citenamefont {Lin}}]{Lin2015}%
  \BibitemOpen
  \bibfield  {author} {\bibinfo {author} {\bibfnamefont {Y.-H.}\ \bibnamefont
  {Lin}}, \bibinfo {author} {\bibfnamefont {S.-F.}\ \bibnamefont {Lin}},
  \bibinfo {author} {\bibfnamefont {Y.-C.}\ \bibnamefont {Chi}}, \bibinfo
  {author} {\bibfnamefont {C.-L.}\ \bibnamefont {Wu}}, \bibinfo {author}
  {\bibfnamefont {C.-H.}\ \bibnamefont {Cheng}}, \bibinfo {author}
  {\bibfnamefont {W.-H.}\ \bibnamefont {Tseng}}, \bibinfo {author}
  {\bibfnamefont {J.-H.}\ \bibnamefont {He}}, \bibinfo {author} {\bibfnamefont
  {C.-I.}\ \bibnamefont {Wu}}, \bibinfo {author} {\bibfnamefont {C.-K.}\
  \bibnamefont {Lee}}, \ and\ \bibinfo {author} {\bibfnamefont {G.-R.}\
  \bibnamefont {Lin}},\ }\href {\doibase 10.1021/acsphotonics.5b00031}
  {\bibfield  {journal} {\bibinfo  {journal} {{ACS} Photonics}\ }\textbf
  {\bibinfo {volume} {2}},\ \bibinfo {pages} {481} (\bibinfo {year}
  {2015})}\BibitemShut {NoStop}%
\bibitem [{\citenamefont {He}\ \emph {et~al.}(2019)\citenamefont {He},
  \citenamefont {Sun},\ and\ \citenamefont {He}}]{He2019}%
  \BibitemOpen
  \bibfield  {author} {\bibinfo {author} {\bibfnamefont {M.}~\bibnamefont
  {He}}, \bibinfo {author} {\bibfnamefont {H.}~\bibnamefont {Sun}}, \ and\
  \bibinfo {author} {\bibfnamefont {Q.~L.}\ \bibnamefont {He}},\ }\href
  {\doibase 10.1007/s11467-019-0893-4} {\bibfield  {journal} {\bibinfo
  {journal} {Frontiers of Physics}\ }\textbf {\bibinfo {volume} {14}} (\bibinfo
  {year} {2019}),\ 10.1007/s11467-019-0893-4}\BibitemShut {NoStop}%
\bibitem [{\citenamefont {Pandey}\ \emph {et~al.}(2021)\citenamefont {Pandey},
  \citenamefont {Yadav}, \citenamefont {Kaur}, \citenamefont {Singh},
  \citenamefont {Gupta},\ and\ \citenamefont {Husale}}]{Pandey2021}%
  \BibitemOpen
  \bibfield  {author} {\bibinfo {author} {\bibfnamefont {A.}~\bibnamefont
  {Pandey}}, \bibinfo {author} {\bibfnamefont {R.}~\bibnamefont {Yadav}},
  \bibinfo {author} {\bibfnamefont {M.}~\bibnamefont {Kaur}}, \bibinfo {author}
  {\bibfnamefont {P.}~\bibnamefont {Singh}}, \bibinfo {author} {\bibfnamefont
  {A.}~\bibnamefont {Gupta}}, \ and\ \bibinfo {author} {\bibfnamefont
  {S.}~\bibnamefont {Husale}},\ }\href {\doibase 10.1038/s41598-020-80738-8}
  {\bibfield  {journal} {\bibinfo  {journal} {Scientific Reports}\ }\textbf
  {\bibinfo {volume} {11}} (\bibinfo {year} {2021}),\
  10.1038/s41598-020-80738-8}\BibitemShut {NoStop}%
\bibitem [{\citenamefont {Breunig}\ and\ \citenamefont
  {Ando}(2021)}]{Breunig2021}%
  \BibitemOpen
  \bibfield  {author} {\bibinfo {author} {\bibfnamefont {O.}~\bibnamefont
  {Breunig}}\ and\ \bibinfo {author} {\bibfnamefont {Y.}~\bibnamefont {Ando}},\
  }\href {\doibase 10.1038/s42254-021-00402-6} {\bibfield  {journal} {\bibinfo
  {journal} {Nature Reviews Physics}\ }\textbf {\bibinfo {volume} {4}},\
  \bibinfo {pages} {184} (\bibinfo {year} {2021})}\BibitemShut {NoStop}%
\bibitem [{\citenamefont {Dziom}\ \emph {et~al.}(2017)\citenamefont {Dziom},
  \citenamefont {Shuvaev}, \citenamefont {Pimenov}, \citenamefont {Astakhov},
  \citenamefont {Ames}, \citenamefont {Bendias}, \citenamefont {Böttcher},
  \citenamefont {Tkachov}, \citenamefont {Hankiewicz}, \citenamefont
  {Br\"{u}ne}, \citenamefont {Buhmann},\ and\ \citenamefont
  {Molenkamp}}]{Dziom2017}%
  \BibitemOpen
  \bibfield  {author} {\bibinfo {author} {\bibfnamefont {V.}~\bibnamefont
  {Dziom}}, \bibinfo {author} {\bibfnamefont {A.}~\bibnamefont {Shuvaev}},
  \bibinfo {author} {\bibfnamefont {A.}~\bibnamefont {Pimenov}}, \bibinfo
  {author} {\bibfnamefont {G.~V.}\ \bibnamefont {Astakhov}}, \bibinfo {author}
  {\bibfnamefont {C.}~\bibnamefont {Ames}}, \bibinfo {author} {\bibfnamefont
  {K.}~\bibnamefont {Bendias}}, \bibinfo {author} {\bibfnamefont
  {J.}~\bibnamefont {Böttcher}}, \bibinfo {author} {\bibfnamefont
  {G.}~\bibnamefont {Tkachov}}, \bibinfo {author} {\bibfnamefont {E.~M.}\
  \bibnamefont {Hankiewicz}}, \bibinfo {author} {\bibfnamefont
  {C.}~\bibnamefont {Br\"{u}ne}}, \bibinfo {author} {\bibfnamefont
  {H.}~\bibnamefont {Buhmann}}, \ and\ \bibinfo {author} {\bibfnamefont
  {L.~W.}\ \bibnamefont {Molenkamp}},\ }\href {\doibase 10.1038/ncomms15197}
  {\bibfield  {journal} {\bibinfo  {journal} {Nature Communications}\ }\textbf
  {\bibinfo {volume} {8}} (\bibinfo {year} {2017}),\
  10.1038/ncomms15197}\BibitemShut {NoStop}%
\bibitem [{\citenamefont {C}\ and\ \citenamefont {F}(2017)}]{JC2017}%
  \BibitemOpen
  \bibfield  {author} {\bibinfo {author} {\bibfnamefont {G.~E.~J.}\
  \bibnamefont {C}}\ and\ \bibinfo {author} {\bibfnamefont {R.~D.}\
  \bibnamefont {F}},\ }\href {\doibase 10.1088/1742-6596/850/1/012024}
  {\bibfield  {journal} {\bibinfo  {journal} {Journal of Physics: Conference
  Series}\ }\textbf {\bibinfo {volume} {850}},\ \bibinfo {pages} {012024}
  (\bibinfo {year} {2017})}\BibitemShut {NoStop}%
\bibitem [{\citenamefont {Crosse}(2017)}]{Crosse2017}%
  \BibitemOpen
  \bibfield  {author} {\bibinfo {author} {\bibfnamefont {J.~A.}\ \bibnamefont
  {Crosse}},\ }\href {\doibase 10.1038/srep43115} {\bibfield  {journal}
  {\bibinfo  {journal} {Scientific Reports}\ }\textbf {\bibinfo {volume} {7}}
  (\bibinfo {year} {2017}),\ 10.1038/srep43115}\BibitemShut {NoStop}%
\bibitem [{\citenamefont {Shuvaev}\ \emph {et~al.}(2022)\citenamefont
  {Shuvaev}, \citenamefont {Pan}, \citenamefont {Tai}, \citenamefont {Zhang},
  \citenamefont {Wang},\ and\ \citenamefont {Pimenov}}]{AShuvaev:2022}%
  \BibitemOpen
  \bibfield  {author} {\bibinfo {author} {\bibfnamefont {A.}~\bibnamefont
  {Shuvaev}}, \bibinfo {author} {\bibfnamefont {L.}~\bibnamefont {Pan}},
  \bibinfo {author} {\bibfnamefont {L.}~\bibnamefont {Tai}}, \bibinfo {author}
  {\bibfnamefont {P.}~\bibnamefont {Zhang}}, \bibinfo {author} {\bibfnamefont
  {K.~L.}\ \bibnamefont {Wang}}, \ and\ \bibinfo {author} {\bibfnamefont
  {A.}~\bibnamefont {Pimenov}},\ }\href {\doibase 10.1063/5.0105159} {\bibfield
   {journal} {\bibinfo  {journal} {Applied Physics Letters}\ }\textbf {\bibinfo
  {volume} {121}},\ \bibinfo {pages} {193101} (\bibinfo {year}
  {2022})}\BibitemShut {NoStop}%
\bibitem [{\citenamefont {Tse}\ and\ \citenamefont
  {MacDonald}(2010)}]{Tse_2010}%
  \BibitemOpen
  \bibfield  {author} {\bibinfo {author} {\bibfnamefont {W.-K.}\ \bibnamefont
  {Tse}}\ and\ \bibinfo {author} {\bibfnamefont {A.~H.}\ \bibnamefont
  {MacDonald}},\ }\href {\doibase 10.1103/physrevlett.105.057401} {\bibfield
  {journal} {\bibinfo  {journal} {Physical Review Letters}\ }\textbf {\bibinfo
  {volume} {105}} (\bibinfo {year} {2010}),\
  10.1103/physrevlett.105.057401}\BibitemShut {NoStop}%
\bibitem [{\citenamefont {Tse}\ and\ \citenamefont
  {MacDonald}(2011)}]{Tse_2011}%
  \BibitemOpen
  \bibfield  {author} {\bibinfo {author} {\bibfnamefont {W.-K.}\ \bibnamefont
  {Tse}}\ and\ \bibinfo {author} {\bibfnamefont {A.~H.}\ \bibnamefont
  {MacDonald}},\ }\href {\doibase 10.1103/physrevb.84.205327} {\bibfield
  {journal} {\bibinfo  {journal} {Physical Review B}\ }\textbf {\bibinfo
  {volume} {84}} (\bibinfo {year} {2011}),\
  10.1103/physrevb.84.205327}\BibitemShut {NoStop}%
\bibitem [{\citenamefont {Crosse}(2016)}]{Crosse_2016}%
  \BibitemOpen
  \bibfield  {author} {\bibinfo {author} {\bibfnamefont {J.~A.}\ \bibnamefont
  {Crosse}},\ }\href {\doibase 10.1103/physreva.94.033816} {\bibfield
  {journal} {\bibinfo  {journal} {Physical Review A}\ }\textbf {\bibinfo
  {volume} {94}} (\bibinfo {year} {2016}),\
  10.1103/physreva.94.033816}\BibitemShut {NoStop}%
\bibitem [{\citenamefont {Ardakani}\ and\ \citenamefont
  {Zare}(2021)}]{GhasempourArdakani:21}%
  \BibitemOpen
  \bibfield  {author} {\bibinfo {author} {\bibfnamefont {A.~G.}\ \bibnamefont
  {Ardakani}}\ and\ \bibinfo {author} {\bibfnamefont {Z.}~\bibnamefont
  {Zare}},\ }\href {\doibase 10.1364/JOSAB.431370} {\bibfield  {journal}
  {\bibinfo  {journal} {J. Opt. Soc. Am. B}\ }\textbf {\bibinfo {volume}
  {38}},\ \bibinfo {pages} {2562} (\bibinfo {year} {2021})}\BibitemShut
  {NoStop}%
\bibitem [{\citenamefont {Mart{\'{\i}}n-Ruiz}\ \emph
  {et~al.}(2016)\citenamefont {Mart{\'{\i}}n-Ruiz}, \citenamefont {Cambiaso},\
  and\ \citenamefont {Urrutia}}]{MartnRuiz2016}%
  \BibitemOpen
  \bibfield  {author} {\bibinfo {author} {\bibfnamefont {A.}~\bibnamefont
  {Mart{\'{\i}}n-Ruiz}}, \bibinfo {author} {\bibfnamefont {M.}~\bibnamefont
  {Cambiaso}}, \ and\ \bibinfo {author} {\bibfnamefont {L.}~\bibnamefont
  {Urrutia}},\ }\href {\doibase 10.1103/physrevd.94.085019} {\bibfield
  {journal} {\bibinfo  {journal} {Physical Review D}\ }\textbf {\bibinfo
  {volume} {94}} (\bibinfo {year} {2016}),\
  10.1103/physrevd.94.085019}\BibitemShut {NoStop}%
\bibitem [{\citenamefont {Ge}\ \emph {et~al.}(2014)\citenamefont {Ge},
  \citenamefont {Zhan}, \citenamefont {Han}, \citenamefont {Liu},\ and\
  \citenamefont {Zi}}]{Ge2014}%
  \BibitemOpen
  \bibfield  {author} {\bibinfo {author} {\bibfnamefont {L.}~\bibnamefont
  {Ge}}, \bibinfo {author} {\bibfnamefont {T.}~\bibnamefont {Zhan}}, \bibinfo
  {author} {\bibfnamefont {D.}~\bibnamefont {Han}}, \bibinfo {author}
  {\bibfnamefont {X.}~\bibnamefont {Liu}}, \ and\ \bibinfo {author}
  {\bibfnamefont {J.}~\bibnamefont {Zi}},\ }\href {\doibase
  10.1364/oe.22.030833} {\bibfield  {journal} {\bibinfo  {journal} {Optics
  Express}\ }\textbf {\bibinfo {volume} {22}},\ \bibinfo {pages} {30833}
  (\bibinfo {year} {2014})}\BibitemShut {NoStop}%
\bibitem [{\citenamefont {Gangaraj}\ \emph {et~al.}(2020)\citenamefont
  {Gangaraj}, \citenamefont {Valagiannopoulos},\ and\ \citenamefont
  {Monticone}}]{HassaniGangaraj2020}%
  \BibitemOpen
  \bibfield  {author} {\bibinfo {author} {\bibfnamefont {S.~A.~H.}\
  \bibnamefont {Gangaraj}}, \bibinfo {author} {\bibfnamefont {C.}~\bibnamefont
  {Valagiannopoulos}}, \ and\ \bibinfo {author} {\bibfnamefont
  {F.}~\bibnamefont {Monticone}},\ }\href {\doibase
  10.1103/physrevresearch.2.023180} {\bibfield  {journal} {\bibinfo  {journal}
  {Physical Review Research}\ }\textbf {\bibinfo {volume} {2}} (\bibinfo {year}
  {2020}),\ 10.1103/physrevresearch.2.023180}\BibitemShut {NoStop}%
\bibitem [{\citenamefont {Mogi}\ \emph {et~al.}(2022)\citenamefont {Mogi},
  \citenamefont {Okamura}, \citenamefont {Kawamura}, \citenamefont {Yoshimi},
  \citenamefont {Yasuda}, \citenamefont {Tsukazaki}, \citenamefont {Takahashi},
  \citenamefont {Morimoto}, \citenamefont {Nagaosa}, \citenamefont {Kawasaki},
  \citenamefont {Takahashi},\ and\ \citenamefont {Tokura}}]{Mogi2022}%
  \BibitemOpen
  \bibfield  {author} {\bibinfo {author} {\bibfnamefont {M.}~\bibnamefont
  {Mogi}}, \bibinfo {author} {\bibfnamefont {Y.}~\bibnamefont {Okamura}},
  \bibinfo {author} {\bibfnamefont {M.}~\bibnamefont {Kawamura}}, \bibinfo
  {author} {\bibfnamefont {R.}~\bibnamefont {Yoshimi}}, \bibinfo {author}
  {\bibfnamefont {K.}~\bibnamefont {Yasuda}}, \bibinfo {author} {\bibfnamefont
  {A.}~\bibnamefont {Tsukazaki}}, \bibinfo {author} {\bibfnamefont {K.~S.}\
  \bibnamefont {Takahashi}}, \bibinfo {author} {\bibfnamefont {T.}~\bibnamefont
  {Morimoto}}, \bibinfo {author} {\bibfnamefont {N.}~\bibnamefont {Nagaosa}},
  \bibinfo {author} {\bibfnamefont {M.}~\bibnamefont {Kawasaki}}, \bibinfo
  {author} {\bibfnamefont {Y.}~\bibnamefont {Takahashi}}, \ and\ \bibinfo
  {author} {\bibfnamefont {Y.}~\bibnamefont {Tokura}},\ }\href {\doibase
  10.1038/s41567-021-01490-y} {\bibfield  {journal} {\bibinfo  {journal}
  {Nature Physics}\ }\textbf {\bibinfo {volume} {18}},\ \bibinfo {pages} {390}
  (\bibinfo {year} {2022})}\BibitemShut {NoStop}%
\bibitem [{\citenamefont {Mart{\'{\i}}n-Ruiz}\ \emph
  {et~al.}(2021)\citenamefont {Mart{\'{\i}}n-Ruiz}, \citenamefont {Cambiaso},\
  and\ \citenamefont {Urrutia}}]{MartnRuiz2021}%
  \BibitemOpen
  \bibfield  {author} {\bibinfo {author} {\bibfnamefont {A.}~\bibnamefont
  {Mart{\'{\i}}n-Ruiz}}, \bibinfo {author} {\bibfnamefont {M.}~\bibnamefont
  {Cambiaso}}, \ and\ \bibinfo {author} {\bibfnamefont {L.~F.}\ \bibnamefont
  {Urrutia}},\ }in\ \href {\doibase 10.1007/978-3-030-62844-4_17} {\emph
  {\bibinfo {booktitle} {Topics in Applied Physics}}}\ (\bibinfo  {publisher}
  {Springer International Publishing},\ \bibinfo {year} {2021})\ pp.\ \bibinfo
  {pages} {459--492}\BibitemShut {NoStop}%
\bibitem [{\citenamefont {Crosse}\ \emph {et~al.}(2015)\citenamefont {Crosse},
  \citenamefont {Fuchs},\ and\ \citenamefont {Buhmann}}]{TSB-TI}%
  \BibitemOpen
  \bibfield  {author} {\bibinfo {author} {\bibfnamefont {J.~A.}\ \bibnamefont
  {Crosse}}, \bibinfo {author} {\bibfnamefont {S.}~\bibnamefont {Fuchs}}, \
  and\ \bibinfo {author} {\bibfnamefont {S.~Y.}\ \bibnamefont {Buhmann}},\
  }\href {\doibase 10.1103/PhysRevA.92.063831} {\bibfield  {journal} {\bibinfo
  {journal} {Phys. Rev. A}\ }\textbf {\bibinfo {volume} {92}},\ \bibinfo
  {pages} {063831} (\bibinfo {year} {2015})}\BibitemShut {NoStop}%
\bibitem [{\citenamefont {Chew}(1995)}]{weng_cho_waves}%
  \BibitemOpen
  \bibfield  {author} {\bibinfo {author} {\bibfnamefont {W.~C.}\ \bibnamefont
  {Chew}},\ }\href@noop {} {\emph {\bibinfo {title} {Waves and fields in
  inhomogeneous media}}}\ (\bibinfo  {publisher} {IEEE Press},\ \bibinfo
  {address} {New York},\ \bibinfo {year} {1995})\BibitemShut {NoStop}%
\bibitem [{\citenamefont {Novotny}\ and\ \citenamefont
  {Hecht}(2012)}]{nano_optics_2012}%
  \BibitemOpen
  \bibfield  {author} {\bibinfo {author} {\bibfnamefont {L.}~\bibnamefont
  {Novotny}}\ and\ \bibinfo {author} {\bibfnamefont {B.}~\bibnamefont
  {Hecht}},\ }\href {\doibase 10.1017/CBO9780511794193} {\emph {\bibinfo
  {title} {Principles of Nano-Optics}}},\ \bibinfo {edition} {2nd}\ ed.\
  (\bibinfo  {publisher} {Cambridge University Press},\ \bibinfo {year}
  {2012})\BibitemShut {NoStop}%
\bibitem [{\citenamefont {Liu}\ and\ \citenamefont {Hesjedal}(2021)}]{Liu2021}%
  \BibitemOpen
  \bibfield  {author} {\bibinfo {author} {\bibfnamefont {J.}~\bibnamefont
  {Liu}}\ and\ \bibinfo {author} {\bibfnamefont {T.}~\bibnamefont {Hesjedal}},\
  }\href {\doibase 10.1002/adma.202102427} {\bibfield  {journal} {\bibinfo
  {journal} {Advanced Materials}\ ,\ \bibinfo {pages} {2102427}} (\bibinfo
  {year} {2021})}\BibitemShut {NoStop}%
\bibitem [{\citenamefont {Mogi}\ \emph {et~al.}(2017)\citenamefont {Mogi},
  \citenamefont {Kawamura}, \citenamefont {Yoshimi}, \citenamefont {Tsukazaki},
  \citenamefont {Kozuka}, \citenamefont {Shirakawa}, \citenamefont {Takahashi},
  \citenamefont {Kawasaki},\ and\ \citenamefont {Tokura}}]{Mogi2017}%
  \BibitemOpen
  \bibfield  {author} {\bibinfo {author} {\bibfnamefont {M.}~\bibnamefont
  {Mogi}}, \bibinfo {author} {\bibfnamefont {M.}~\bibnamefont {Kawamura}},
  \bibinfo {author} {\bibfnamefont {R.}~\bibnamefont {Yoshimi}}, \bibinfo
  {author} {\bibfnamefont {A.}~\bibnamefont {Tsukazaki}}, \bibinfo {author}
  {\bibfnamefont {Y.}~\bibnamefont {Kozuka}}, \bibinfo {author} {\bibfnamefont
  {N.}~\bibnamefont {Shirakawa}}, \bibinfo {author} {\bibfnamefont {K.~S.}\
  \bibnamefont {Takahashi}}, \bibinfo {author} {\bibfnamefont {M.}~\bibnamefont
  {Kawasaki}}, \ and\ \bibinfo {author} {\bibfnamefont {Y.}~\bibnamefont
  {Tokura}},\ }\href {\doibase 10.1038/nmat4855} {\bibfield  {journal}
  {\bibinfo  {journal} {Nature Materials}\ }\textbf {\bibinfo {volume} {16}},\
  \bibinfo {pages} {516} (\bibinfo {year} {2017})}\BibitemShut {NoStop}%
\bibitem [{\citenamefont {An}\ \emph {et~al.}(2020)\citenamefont {An},
  \citenamefont {Zeng}, \citenamefont {Zhang}, \citenamefont {Li},
  \citenamefont {Hu}, \citenamefont {Li},\ and\ \citenamefont {Zeng}}]{An2020}%
  \BibitemOpen
  \bibfield  {author} {\bibinfo {author} {\bibfnamefont {H.}~\bibnamefont
  {An}}, \bibinfo {author} {\bibfnamefont {R.}~\bibnamefont {Zeng}}, \bibinfo
  {author} {\bibfnamefont {M.}~\bibnamefont {Zhang}}, \bibinfo {author}
  {\bibfnamefont {H.}~\bibnamefont {Li}}, \bibinfo {author} {\bibfnamefont
  {M.}~\bibnamefont {Hu}}, \bibinfo {author} {\bibfnamefont {Q.}~\bibnamefont
  {Li}}, \ and\ \bibinfo {author} {\bibfnamefont {X.}~\bibnamefont {Zeng}},\
  }\href {\doibase 10.1016/j.optcom.2020.126335} {\bibfield  {journal}
  {\bibinfo  {journal} {Optics Communications}\ }\textbf {\bibinfo {volume}
  {477}},\ \bibinfo {pages} {126335} (\bibinfo {year} {2020})}\BibitemShut
  {NoStop}%
\bibitem [{\citenamefont {Zuo}\ \emph {et~al.}(2013)\citenamefont {Zuo},
  \citenamefont {Ling}, \citenamefont {Sheng},\ and\ \citenamefont
  {Xing}}]{Zuo2013}%
  \BibitemOpen
  \bibfield  {author} {\bibinfo {author} {\bibfnamefont {Z.-W.}\ \bibnamefont
  {Zuo}}, \bibinfo {author} {\bibfnamefont {D.-B.}\ \bibnamefont {Ling}},
  \bibinfo {author} {\bibfnamefont {L.}~\bibnamefont {Sheng}}, \ and\ \bibinfo
  {author} {\bibfnamefont {D.}~\bibnamefont {Xing}},\ }\href {\doibase
  10.1016/j.physleta.2013.09.004} {\bibfield  {journal} {\bibinfo  {journal}
  {Physics Letters A}\ }\textbf {\bibinfo {volume} {377}},\ \bibinfo {pages}
  {2909} (\bibinfo {year} {2013})}\BibitemShut {NoStop}%
\bibitem [{\citenamefont {Sato}\ \emph {et~al.}(2010)\citenamefont {Sato},
  \citenamefont {Segawa}, \citenamefont {Guo}, \citenamefont {Sugawara},
  \citenamefont {Souma}, \citenamefont {Takahashi},\ and\ \citenamefont
  {Ando}}]{PhysRevLett.105.136802}%
  \BibitemOpen
  \bibfield  {author} {\bibinfo {author} {\bibfnamefont {T.}~\bibnamefont
  {Sato}}, \bibinfo {author} {\bibfnamefont {K.}~\bibnamefont {Segawa}},
  \bibinfo {author} {\bibfnamefont {H.}~\bibnamefont {Guo}}, \bibinfo {author}
  {\bibfnamefont {K.}~\bibnamefont {Sugawara}}, \bibinfo {author}
  {\bibfnamefont {S.}~\bibnamefont {Souma}}, \bibinfo {author} {\bibfnamefont
  {T.}~\bibnamefont {Takahashi}}, \ and\ \bibinfo {author} {\bibfnamefont
  {Y.}~\bibnamefont {Ando}},\ }\href {\doibase 10.1103/PhysRevLett.105.136802}
  {\bibfield  {journal} {\bibinfo  {journal} {Phys. Rev. Lett.}\ }\textbf
  {\bibinfo {volume} {105}},\ \bibinfo {pages} {136802} (\bibinfo {year}
  {2010})}\BibitemShut {NoStop}%
\bibitem [{\citenamefont {Mitsas}\ \emph {et~al.}()\citenamefont {Mitsas},
  \citenamefont {Siapkas}, \citenamefont {Polychroniadis}, \citenamefont
  {Valassiades},\ and\ \citenamefont {Paraskevopoulos}}]{MitsasGrowth}%
  \BibitemOpen
  \bibfield  {author} {\bibinfo {author} {\bibfnamefont {C.}~\bibnamefont
  {Mitsas}}, \bibinfo {author} {\bibfnamefont {D.~I.}\ \bibnamefont {Siapkas}},
  \bibinfo {author} {\bibfnamefont {E.~K.}\ \bibnamefont {Polychroniadis}},
  \bibinfo {author} {\bibfnamefont {O.}~\bibnamefont {Valassiades}}, \ and\
  \bibinfo {author} {\bibfnamefont {K.~M.}\ \bibnamefont {Paraskevopoulos}},\
  }\href@noop {} {\bibfield  {journal} {\bibinfo  {journal} {Physica status
  solidi}\ }\textbf {\bibinfo {volume} {136}},\ \bibinfo {pages}
  {483}}\BibitemShut {NoStop}%
\bibitem [{\citenamefont {Castro-Enriquez}\ \emph {et~al.}(2020)\citenamefont
  {Castro-Enriquez}, \citenamefont {Quezada},\ and\ \citenamefont
  {Mart\'{\i}n-Ruiz}}]{Castro2020}%
  \BibitemOpen
  \bibfield  {author} {\bibinfo {author} {\bibfnamefont {L.~A.}\ \bibnamefont
  {Castro-Enriquez}}, \bibinfo {author} {\bibfnamefont {L.~F.}\ \bibnamefont
  {Quezada}}, \ and\ \bibinfo {author} {\bibfnamefont {A.}~\bibnamefont
  {Mart\'{\i}n-Ruiz}},\ }\href {\doibase 10.1103/PhysRevA.102.013720}
  {\bibfield  {journal} {\bibinfo  {journal} {Phys. Rev. A}\ }\textbf {\bibinfo
  {volume} {102}},\ \bibinfo {pages} {013720} (\bibinfo {year}
  {2020})}\BibitemShut {NoStop}%
\bibitem [{\citenamefont {Phutela}\ \emph {et~al.}(2022)\citenamefont
  {Phutela}, \citenamefont {Bhumla}, \citenamefont {Jain},\ and\ \citenamefont
  {Bhattacharya}}]{Phutela2022}%
  \BibitemOpen
  \bibfield  {author} {\bibinfo {author} {\bibfnamefont {A.}~\bibnamefont
  {Phutela}}, \bibinfo {author} {\bibfnamefont {P.}~\bibnamefont {Bhumla}},
  \bibinfo {author} {\bibfnamefont {M.}~\bibnamefont {Jain}}, \ and\ \bibinfo
  {author} {\bibfnamefont {S.}~\bibnamefont {Bhattacharya}},\ }\href {\doibase
  10.1038/s41598-022-26445-y} {\bibfield  {journal} {\bibinfo  {journal}
  {Scientific Reports}\ }\textbf {\bibinfo {volume} {12}} (\bibinfo {year}
  {2022}),\ 10.1038/s41598-022-26445-y}\BibitemShut {NoStop}%
\bibitem [{\citenamefont {Chen}\ \emph {et~al.}(2019)\citenamefont {Chen},
  \citenamefont {Fei}, \citenamefont {Zhang}, \citenamefont {Zhang},
  \citenamefont {Liu}, \citenamefont {Zhang}, \citenamefont {Wang},
  \citenamefont {Wei}, \citenamefont {Zhang}, \citenamefont {Zuo},
  \citenamefont {Guo}, \citenamefont {Liu}, \citenamefont {Wang}, \citenamefont
  {Wu}, \citenamefont {Zong}, \citenamefont {Xie}, \citenamefont {Chen},
  \citenamefont {Sun}, \citenamefont {Wang}, \citenamefont {Zhang},
  \citenamefont {Zhang}, \citenamefont {Wang}, \citenamefont {Song},
  \citenamefont {Zhang}, \citenamefont {Shen},\ and\ \citenamefont
  {Wang}}]{Chen2019}%
  \BibitemOpen
  \bibfield  {author} {\bibinfo {author} {\bibfnamefont {B.}~\bibnamefont
  {Chen}}, \bibinfo {author} {\bibfnamefont {F.}~\bibnamefont {Fei}}, \bibinfo
  {author} {\bibfnamefont {D.}~\bibnamefont {Zhang}}, \bibinfo {author}
  {\bibfnamefont {B.}~\bibnamefont {Zhang}}, \bibinfo {author} {\bibfnamefont
  {W.}~\bibnamefont {Liu}}, \bibinfo {author} {\bibfnamefont {S.}~\bibnamefont
  {Zhang}}, \bibinfo {author} {\bibfnamefont {P.}~\bibnamefont {Wang}},
  \bibinfo {author} {\bibfnamefont {B.}~\bibnamefont {Wei}}, \bibinfo {author}
  {\bibfnamefont {Y.}~\bibnamefont {Zhang}}, \bibinfo {author} {\bibfnamefont
  {Z.}~\bibnamefont {Zuo}}, \bibinfo {author} {\bibfnamefont {J.}~\bibnamefont
  {Guo}}, \bibinfo {author} {\bibfnamefont {Q.}~\bibnamefont {Liu}}, \bibinfo
  {author} {\bibfnamefont {Z.}~\bibnamefont {Wang}}, \bibinfo {author}
  {\bibfnamefont {X.}~\bibnamefont {Wu}}, \bibinfo {author} {\bibfnamefont
  {J.}~\bibnamefont {Zong}}, \bibinfo {author} {\bibfnamefont {X.}~\bibnamefont
  {Xie}}, \bibinfo {author} {\bibfnamefont {W.}~\bibnamefont {Chen}}, \bibinfo
  {author} {\bibfnamefont {Z.}~\bibnamefont {Sun}}, \bibinfo {author}
  {\bibfnamefont {S.}~\bibnamefont {Wang}}, \bibinfo {author} {\bibfnamefont
  {Y.}~\bibnamefont {Zhang}}, \bibinfo {author} {\bibfnamefont
  {M.}~\bibnamefont {Zhang}}, \bibinfo {author} {\bibfnamefont
  {X.}~\bibnamefont {Wang}}, \bibinfo {author} {\bibfnamefont {F.}~\bibnamefont
  {Song}}, \bibinfo {author} {\bibfnamefont {H.}~\bibnamefont {Zhang}},
  \bibinfo {author} {\bibfnamefont {D.}~\bibnamefont {Shen}}, \ and\ \bibinfo
  {author} {\bibfnamefont {B.}~\bibnamefont {Wang}},\ }\href {\doibase
  10.1038/s41467-019-12485-y} {\bibfield  {journal} {\bibinfo  {journal}
  {Nature Communications}\ }\textbf {\bibinfo {volume} {10}} (\bibinfo {year}
  {2019}),\ 10.1038/s41467-019-12485-y}\BibitemShut {NoStop}%
\bibitem [{\citenamefont {Kim}\ \emph {et~al.}(2019)\citenamefont {Kim},
  \citenamefont {Shin}, \citenamefont {Sharma}, \citenamefont {Ihm},
  \citenamefont {Dugerjav}, \citenamefont {Hwang}, \citenamefont {Lee},
  \citenamefont {Ko}, \citenamefont {Park}, \citenamefont {Kim}, \citenamefont
  {Kim},\ and\ \citenamefont {Jung}}]{Kim2019}%
  \BibitemOpen
  \bibfield  {author} {\bibinfo {author} {\bibfnamefont {J.}~\bibnamefont
  {Kim}}, \bibinfo {author} {\bibfnamefont {E.-H.}\ \bibnamefont {Shin}},
  \bibinfo {author} {\bibfnamefont {M.~K.}\ \bibnamefont {Sharma}}, \bibinfo
  {author} {\bibfnamefont {K.}~\bibnamefont {Ihm}}, \bibinfo {author}
  {\bibfnamefont {O.}~\bibnamefont {Dugerjav}}, \bibinfo {author}
  {\bibfnamefont {C.}~\bibnamefont {Hwang}}, \bibinfo {author} {\bibfnamefont
  {H.}~\bibnamefont {Lee}}, \bibinfo {author} {\bibfnamefont {K.-T.}\
  \bibnamefont {Ko}}, \bibinfo {author} {\bibfnamefont {J.-H.}\ \bibnamefont
  {Park}}, \bibinfo {author} {\bibfnamefont {M.}~\bibnamefont {Kim}}, \bibinfo
  {author} {\bibfnamefont {H.}~\bibnamefont {Kim}}, \ and\ \bibinfo {author}
  {\bibfnamefont {M.-H.}\ \bibnamefont {Jung}},\ }\href {\doibase
  10.1038/s41598-018-37663-8} {\bibfield  {journal} {\bibinfo  {journal}
  {Scientific Reports}\ }\textbf {\bibinfo {volume} {9}} (\bibinfo {year}
  {2019}),\ 10.1038/s41598-018-37663-8}\BibitemShut {NoStop}%
\bibitem [{\citenamefont {Choi}\ \emph {et~al.}(2011)\citenamefont {Choi},
  \citenamefont {Jo}, \citenamefont {Lee}, \citenamefont {Yoon}, \citenamefont
  {You},\ and\ \citenamefont {Jung}}]{Choi2011}%
  \BibitemOpen
  \bibfield  {author} {\bibinfo {author} {\bibfnamefont {Y.~H.}\ \bibnamefont
  {Choi}}, \bibinfo {author} {\bibfnamefont {N.~H.}\ \bibnamefont {Jo}},
  \bibinfo {author} {\bibfnamefont {K.~J.}\ \bibnamefont {Lee}}, \bibinfo
  {author} {\bibfnamefont {J.~B.}\ \bibnamefont {Yoon}}, \bibinfo {author}
  {\bibfnamefont {C.~Y.}\ \bibnamefont {You}}, \ and\ \bibinfo {author}
  {\bibfnamefont {M.~H.}\ \bibnamefont {Jung}},\ }\href {\doibase
  10.1063/1.3549553} {\bibfield  {journal} {\bibinfo  {journal} {Journal of
  Applied Physics}\ }\textbf {\bibinfo {volume} {109}},\ \bibinfo {pages}
  {07E312} (\bibinfo {year} {2011})}\BibitemShut {NoStop}%
\bibitem [{\citenamefont {Teng}\ \emph {et~al.}(2019)\citenamefont {Teng},
  \citenamefont {Liu},\ and\ \citenamefont {Li}}]{Teng2019}%
  \BibitemOpen
  \bibfield  {author} {\bibinfo {author} {\bibfnamefont {J.}~\bibnamefont
  {Teng}}, \bibinfo {author} {\bibfnamefont {N.}~\bibnamefont {Liu}}, \ and\
  \bibinfo {author} {\bibfnamefont {Y.}~\bibnamefont {Li}},\ }\href {\doibase
  10.1088/1674-4926/40/8/081507} {\bibfield  {journal} {\bibinfo  {journal}
  {Journal of Semiconductors}\ }\textbf {\bibinfo {volume} {40}},\ \bibinfo
  {pages} {081507} (\bibinfo {year} {2019})}\BibitemShut {NoStop}%
\bibitem [{\citenamefont {Yamazaki}\ \emph {et~al.}(2005)\citenamefont
  {Yamazaki}, \citenamefont {Kato}, \citenamefont {Muramatsu}, \citenamefont
  {Haga}, \citenamefont {Kobayashi}, \citenamefont {Kamata}, \citenamefont
  {Fujiwara},\ and\ \citenamefont {Yamaguchi}}]{yamazaki2005incremental}%
  \BibitemOpen
  \bibfield  {author} {\bibinfo {author} {\bibfnamefont {K.}~\bibnamefont
  {Yamazaki}}, \bibinfo {author} {\bibfnamefont {K.}~\bibnamefont {Kato}},
  \bibinfo {author} {\bibfnamefont {K.}~\bibnamefont {Muramatsu}}, \bibinfo
  {author} {\bibfnamefont {A.}~\bibnamefont {Haga}}, \bibinfo {author}
  {\bibfnamefont {K.}~\bibnamefont {Kobayashi}}, \bibinfo {author}
  {\bibfnamefont {K.}~\bibnamefont {Kamata}}, \bibinfo {author} {\bibfnamefont
  {K.}~\bibnamefont {Fujiwara}}, \ and\ \bibinfo {author} {\bibfnamefont
  {T.}~\bibnamefont {Yamaguchi}},\ }\href@noop {} {\bibfield  {journal}
  {\bibinfo  {journal} {IEEE transactions on magnetics}\ }\textbf {\bibinfo
  {volume} {41}},\ \bibinfo {pages} {4087} (\bibinfo {year}
  {2005})}\BibitemShut {NoStop}%
\bibitem [{\citenamefont {ter Brake}\ \emph {et~al.}(1991)\citenamefont {ter
  Brake}, \citenamefont {Wieringa},\ and\ \citenamefont
  {Rogalla}}]{ter1991improvement}%
  \BibitemOpen
  \bibfield  {author} {\bibinfo {author} {\bibfnamefont {H.~J.}\ \bibnamefont
  {ter Brake}}, \bibinfo {author} {\bibfnamefont {H.}~\bibnamefont {Wieringa}},
  \ and\ \bibinfo {author} {\bibfnamefont {H.}~\bibnamefont {Rogalla}},\
  }\href@noop {} {\bibfield  {journal} {\bibinfo  {journal} {Measurement
  Science and Technology}\ }\textbf {\bibinfo {volume} {2}},\ \bibinfo {pages}
  {596} (\bibinfo {year} {1991})}\BibitemShut {NoStop}%
\bibitem [{\citenamefont {Gao}\ \emph {et~al.}(2023)\citenamefont {Gao},
  \citenamefont {Ma}, \citenamefont {Wang}, \citenamefont {Xu}, \citenamefont
  {Li}, \citenamefont {Dou},\ and\ \citenamefont {Li}}]{gao2023low}%
  \BibitemOpen
  \bibfield  {author} {\bibinfo {author} {\bibfnamefont {Y.}~\bibnamefont
  {Gao}}, \bibinfo {author} {\bibfnamefont {D.}~\bibnamefont {Ma}}, \bibinfo
  {author} {\bibfnamefont {K.}~\bibnamefont {Wang}}, \bibinfo {author}
  {\bibfnamefont {X.}~\bibnamefont {Xu}}, \bibinfo {author} {\bibfnamefont
  {S.}~\bibnamefont {Li}}, \bibinfo {author} {\bibfnamefont {Y.}~\bibnamefont
  {Dou}}, \ and\ \bibinfo {author} {\bibfnamefont {J.}~\bibnamefont {Li}},\
  }\href@noop {} {\bibfield  {journal} {\bibinfo  {journal} {Sensors and
  Actuators A: Physical}\ }\textbf {\bibinfo {volume} {352}},\ \bibinfo {pages}
  {114207} (\bibinfo {year} {2023})}\BibitemShut {NoStop}%
\bibitem [{\citenamefont {Dubbers}(1986)}]{dubbers1986simple}%
  \BibitemOpen
  \bibfield  {author} {\bibinfo {author} {\bibfnamefont {D.}~\bibnamefont
  {Dubbers}},\ }\href@noop {} {\bibfield  {journal} {\bibinfo  {journal}
  {Nuclear Instruments and Methods in Physics Research Section A: Accelerators,
  Spectrometers, Detectors and Associated Equipment}\ }\textbf {\bibinfo
  {volume} {243}},\ \bibinfo {pages} {511} (\bibinfo {year}
  {1986})}\BibitemShut {NoStop}%
\bibitem [{\citenamefont {Gaskill}(1978)}]{Gaskill1978-yg}%
  \BibitemOpen
  \bibfield  {author} {\bibinfo {author} {\bibfnamefont {J.~D.}\ \bibnamefont
  {Gaskill}},\ }\href@noop {} {\emph {\bibinfo {title} {Linear systems, Fourier
  transforms, and optics}}},\ Wiley Series in Pure and Applied Optics\
  (\bibinfo  {publisher} {John Wiley \& Sons},\ \bibinfo {address} {Nashville,
  TN},\ \bibinfo {year} {1978})\BibitemShut {NoStop}%
\end{thebibliography}%

\onecolumngrid
\newpage
\appendix
\section{Green's Function components plots for different configurations}\label{appendix:G_graphs}
\begin{figure}[!htp]
    \centering
    \includegraphics[width=0.68\linewidth]{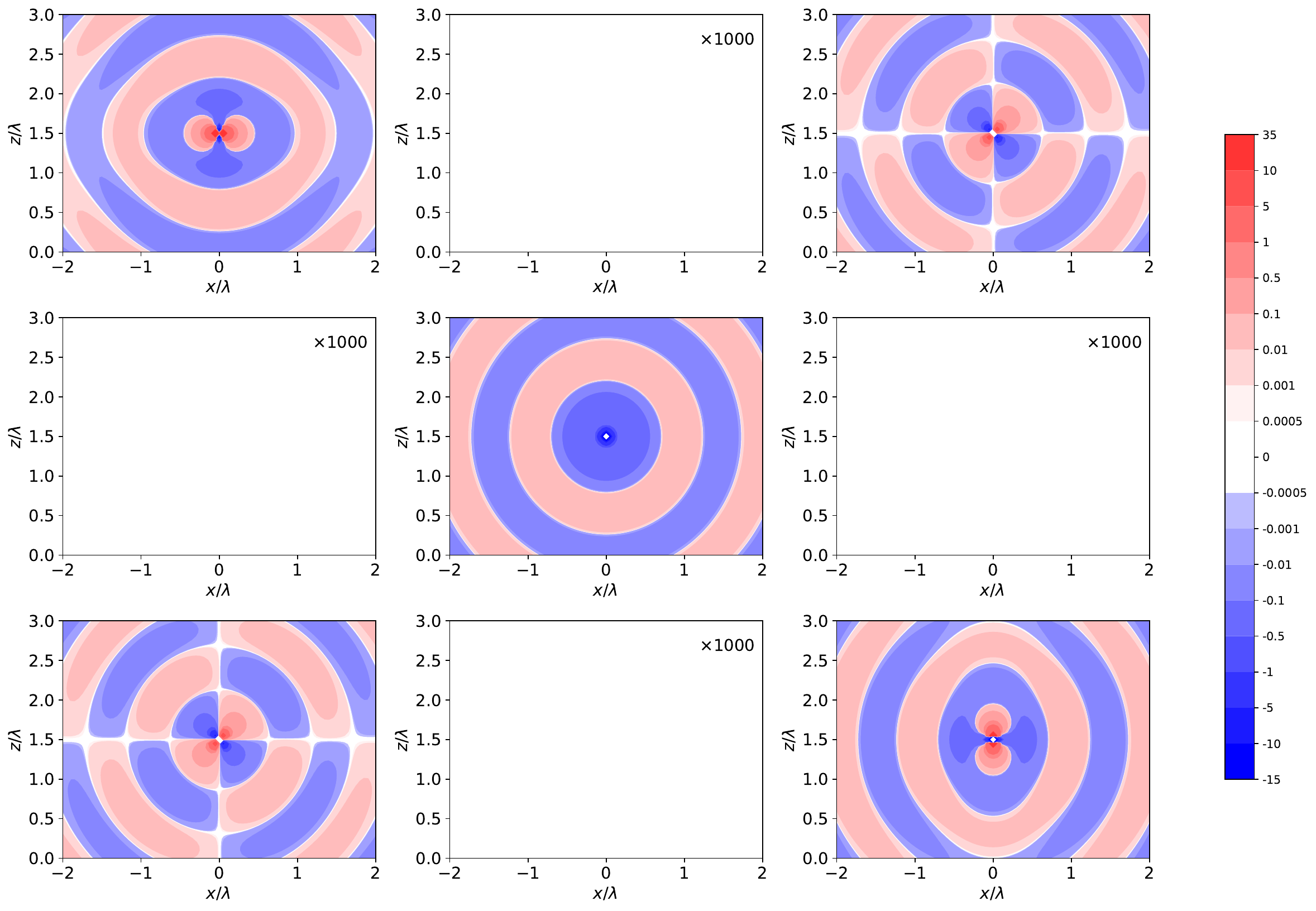}
    \caption{Respective matrix components of $\mathbf{G}_0(\mathbf{r},\mathbf{r}',\omega)\times \lambda$, graphed in the $y=0$ plane, as a function of $z/\lambda$ and $x/\lambda$ for free-space, with parameters $\mu_1=\epsilon_1=1$ and $\lambda=600 \;\text{nm}$. Here, the emitting dipole is arbitrarily located at $\mathbf{r}'=1.5\times\lambda$. Here $\mathbf{G}_{ij}$ can be understood as the $E_i$ electric field component generated by a dipole oriented in $j$. Note that the off-diagonal components are null, regardless of the enhancement factor.}
    \label{fig:G0_plot}
\end{figure}

\begin{figure}[!htp]
    \centering
    \includegraphics[width=0.68\linewidth]{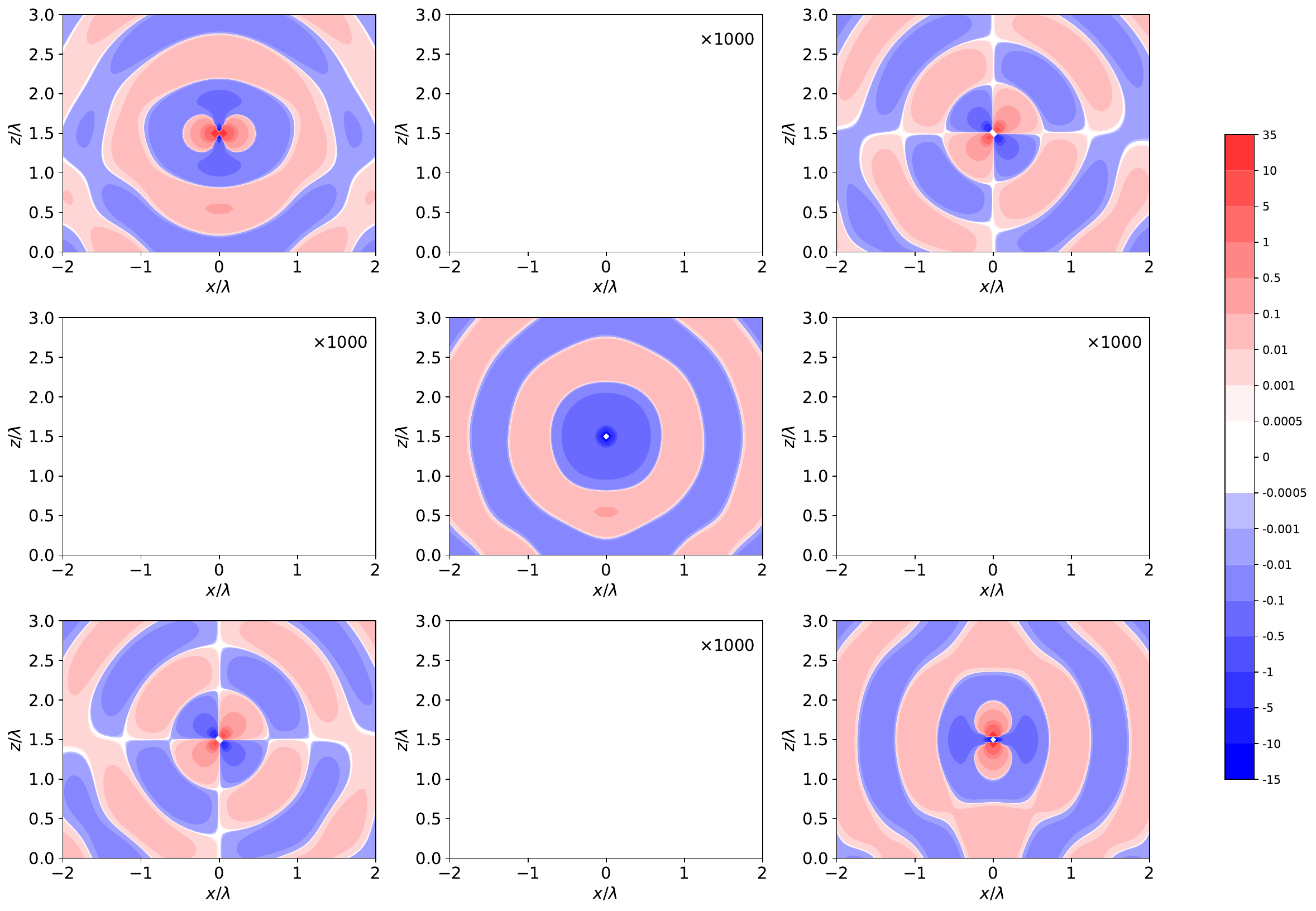}
    \caption{Respective matrix components of $\mathbf{G}(\mathbf{r},\mathbf{r}',\omega)\times \lambda$, graphed in the $y=0$ plane, as a function of $z/\lambda$ and $x/\lambda$ for the Air/Magnetodielectric configuration, with parameters $\mu_1=\epsilon_1=\epsilon_2=1$, $\mu_2=2$, $\Theta_2=0$ and $\lambda=600 \;\text{nm}$, and the interface located at $z=0$. Here, the emitting dipole is arbitrarily located at $\mathbf{r}'=1.5\times\lambda$. Here $\mathbf{G}_{ij}$ can be understood as the $E_i$ electric field component generated by a dipole oriented in $j$. Note that the off-diagonal components remain null, regardless of the enhancement factor.}
    \label{fig:Gmd_plot}
\end{figure}

\begin{figure}[!htp]
    \centering
    \includegraphics[width=0.72\linewidth]{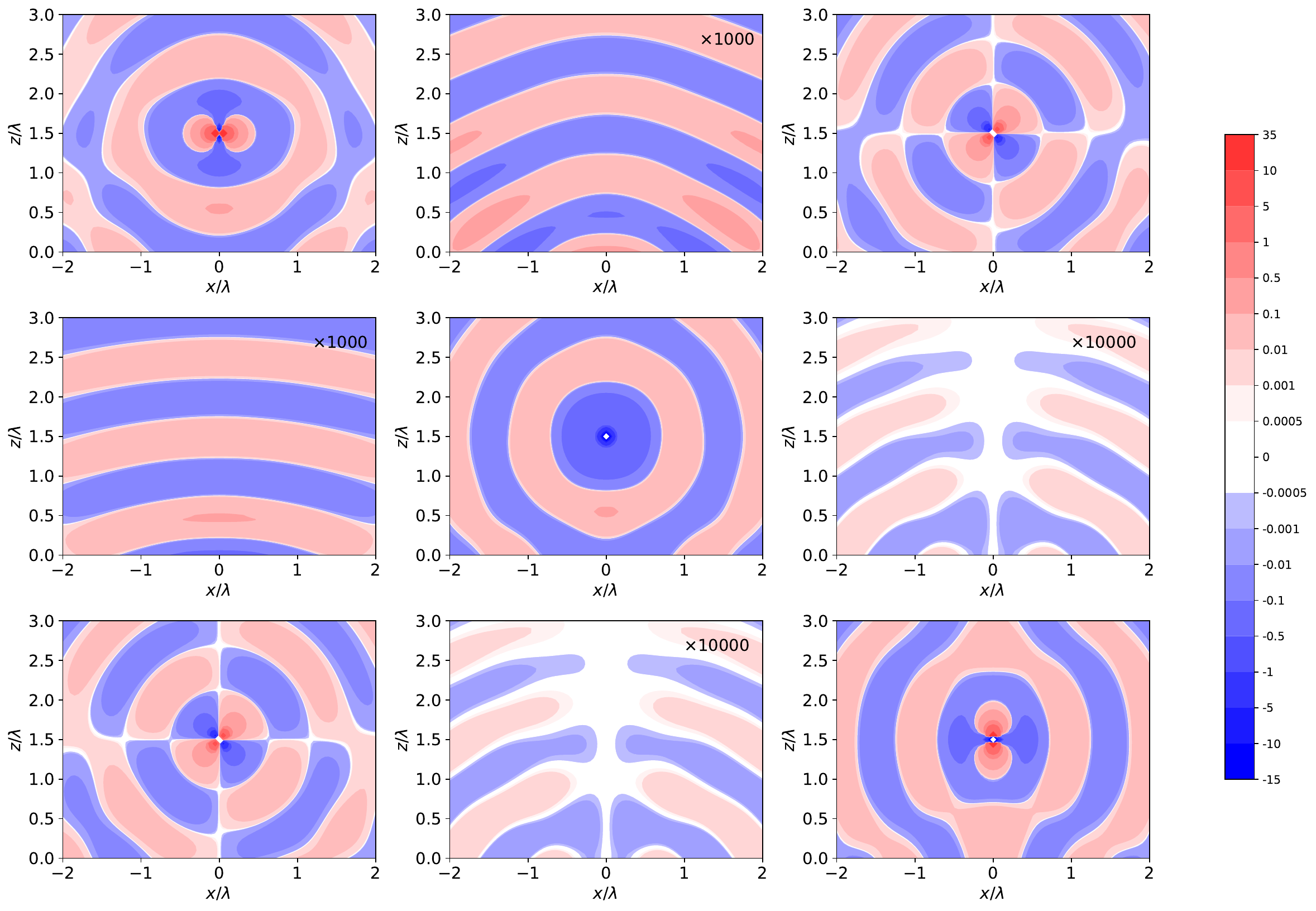}
    \caption{Respective matrix components of $\mathbf{G}(\mathbf{r},\mathbf{r}',\omega)\times \lambda$, graphed in the $y=0$ plane, as a function of $z/\lambda$ and $x/\lambda$ for the Air/TI configuration, with parameters $\mu_1=\epsilon_1=\epsilon_2=1$, $\mu_2=2$, $\Theta_2=\pi$ and $\lambda=600 \;\text{nm}$, and the interface located at $z=0$. Here, the emitting dipole is arbitrarily located at $\mathbf{r}'=1.5\times\lambda$. Here $\mathbf{G}_{ij}$ can be understood as the $E_i$ electric field component generated by a dipole oriented in $j$. Note that here we had to enhance the off-diagonal components by a factor of $1000$ and $10000$, respectively, to see them with the normal emissions. These components appear due to topological contributions.}
    \label{fig:G_2-layered_plot}
\end{figure}

\begin{figure}[!htp]
    \centering
    \includegraphics[width=0.72\linewidth]{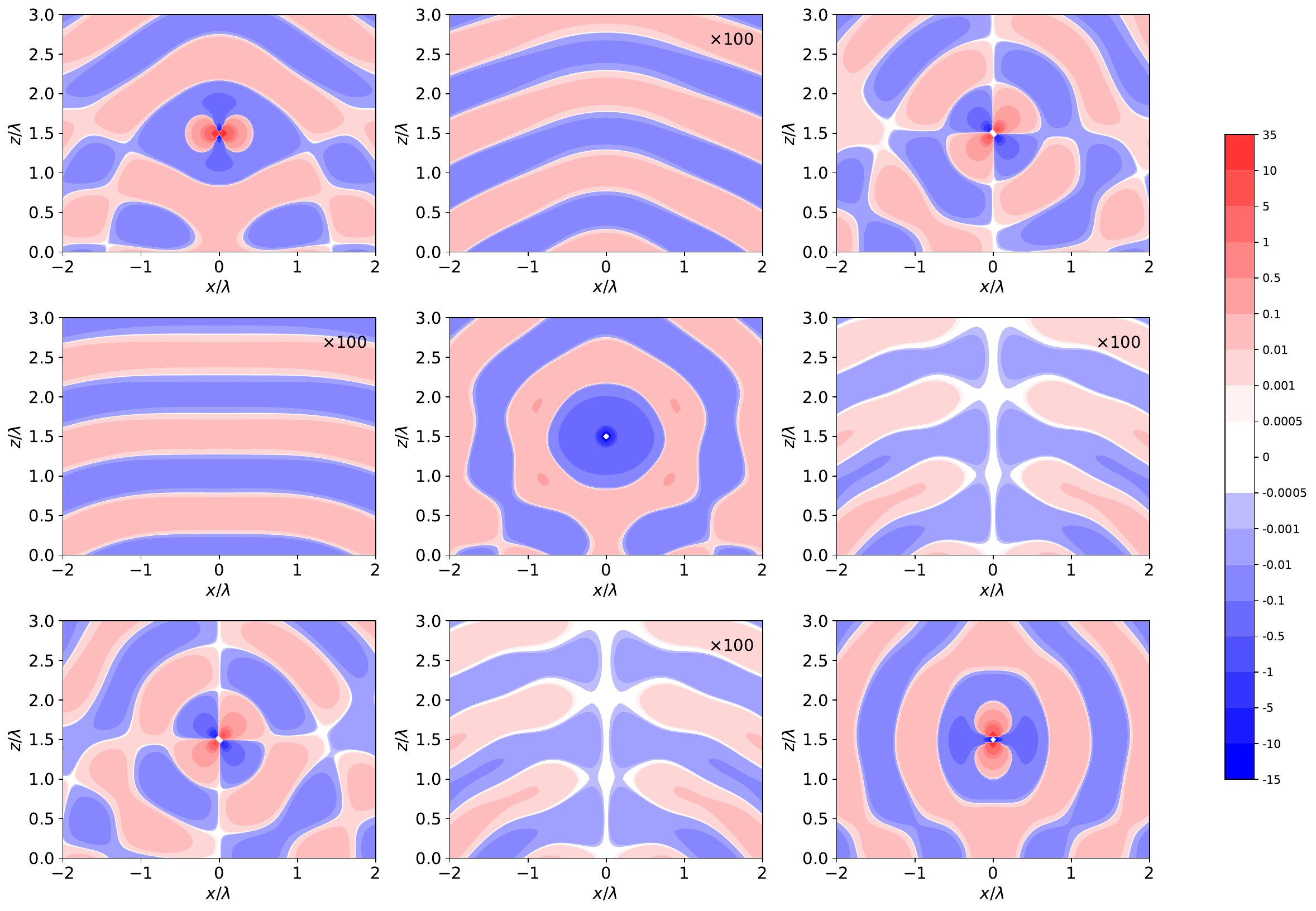}
    \caption{Respective matrix components of $\mathbf{G}(\mathbf{r},\mathbf{r}',\omega)\times \lambda$, graphed in the $y=0$ plane, as a function of $z/\lambda$ and $x/\lambda$ for the Air/TI/Mu-metal configuration, with parameters $\mu_1=\epsilon_1=\epsilon_2=\epsilon_3=1$, $\mu_2=2$, $\mu_3=10^5$, $\Theta_2=\pi$, $\Theta_3=0$, $d=100\;\text{nm}$ (thickness of the TI-layer) and $\lambda=600 \;\text{nm}$. Here, the emitting dipole is arbitrarily located at $\mathbf{r}'=1.5\times\lambda$. Here $\mathbf{G}_{ij}$ can be understood as the $E_i$ electric field component generated by a dipole oriented in $j$. Note that here we had to enhance the off-diagonal components by a factor of $100$ to see them with the normal emissions. This implies an enhancement of an order of magnitude for $G_{xy}$ and $G_{yx}$, and of two orders of magnitude for $G_{yz}$ and $G_{zy}$, compared with the Air/TI configuration.}
    \label{fig:G_3-layered_plot}
\end{figure}

\end{document}